\documentclass[final,5p,times,twocolumn]{elsarticle}

\usepackage{amssymb}

\usepackage{balance}
\biboptions{sort&compress}
\usepackage{graphicx}
\usepackage{picinpar}
\usepackage{amsmath}
\usepackage{url}
\usepackage[latin1]{inputenc}
\usepackage{colortbl}
\usepackage{soul}
\usepackage{multirow}
\usepackage{pifont}
\usepackage{alltt}
\usepackage[hidelinks]{hyperref}
\usepackage{enumerate}
\usepackage{siunitx}
\usepackage{breakurl}
\usepackage{pbox}
\usepackage{tabularx}

\usepackage{amsthm}
\theoremstyle{remark}
\newtheorem{remark}{Remark}
\theoremstyle{plain}
\usepackage[caption=false]{subfig}
\usepackage{amsfonts}
\usepackage{booktabs}
\usepackage{array}
\usepackage{makecell}
\usepackage{empheq}
\usepackage{tikz}
\usetikzlibrary{shapes.geometric, arrows}

\journal{Applied Energy}

\begin{document}

\begin{frontmatter}



\title{ Remaining Discharge Energy Prediction for Lithium-Ion Batteries Over Broad Current Ranges: A Machine Learning Approach}


\author[inst1]{Hao Tu}

\affiliation[inst1]{organization={Department of Mechanical Engineering, University of Kansas},
            city={Lawrence},
            postcode={KS 66045}, 
            country={USA}}

\author[inst4,inst2]{Manashita Borah}

\affiliation[inst4]{organization={Department of Electrical Engineering, Tezpur University},
            city={Tezpur},
            postcode={784028}, 
            country={India}}

\author[inst2]{Scott Moura}

\affiliation[inst2]{organization={Department of Civil and Environmental Engineering, University of California},
            city={Berkeley},
            postcode={CA 94720}, 
            country={USA}}

\author[inst3]{Yebin Wang}

\affiliation[inst3]{organization={Mitsubishi Electric Research Laboratories},
            city={Cambridge},
            postcode={MA 02139}, 
            country={USA}}

\author[inst1]{Huazhen Fang\corref{cor1}}
\ead{fang@ku.edu}
\cortext[cor1]{Corresponding author}

\begin{abstract}
Lithium-ion batteries have found their way into myriad sectors of industry to drive electrification, decarbonization, and sustainability. A crucial aspect in ensuring their safe and optimal performance is monitoring their energy levels. In this paper, we present the first study on predicting the remaining energy of a battery cell undergoing discharge over wide current ranges from low to high C-rates. The complexity of the challenge arises from the cell's C-rate-dependent energy availability as well as its intricate electro-thermal dynamics especially at high C-rates. To address this, we introduce a new definition of remaining discharge energy and then undertake a systematic effort in harnessing the power of machine learning to enable its prediction. Our effort includes two parts in cascade. First, we develop an accurate dynamic model based on integration of physics with machine learning to capture a battery's voltage and temperature behaviors. Second, based on the model, we propose a machine learning approach to predict the remaining discharge energy under arbitrary C-rates and pre-specified cut-off limits in voltage and temperature. The experimental validation shows that the proposed approach can predict the remaining discharge energy with a relative error of less than 3\% when the current varies between 0$\sim$8 C for an NCA cell and 0$\sim$15 C for an LFP cell. The approach, by design, is amenable to training and computation.
\end{abstract}



\begin{keyword}
Lithium-ion batteries \sep machine learning \sep neural networks \sep remaining discharge energy 


\end{keyword}

\end{frontmatter}


\section{Introduction}
\label{sec: Introduction}

Lithium-ion batteries (LiBs) represent one of the most important power source technologies of our time. They have transformed the consumer electronics sector since the 1990s and are now driving the revolution of transportation electrification that extends from passenger cars to commercial vehicles to aircraft.  Battery management systems (BMSs) must be in place to ensure LiBs' operational safety and performance. Among the various BMS functions, a significant one is monitoring the state of energy or, more specifically, the remaining discharge energy (RDE). While its definition may take different forms, the RDE generally indicates how much energy is available before the cell gets depleted in discharging. Practical applications demand accurate, real-time RDE prediction to avoid over-discharging or to determine the remaining duration or range in power supply. However, the prediction is non-trivial due to the complex dynamics of LiBs, which has attracted a growing body of research in recent years.

Closely related to the RDE, a more familiar quantity in the literature is the so-called state-of-charge (SoC), which is the percentage ratio between a cell's available charge capacity in ampere-hours (Ah) and its nominal capacity. While the estimation of it has received extensive study~\cite{Wang:ICSM:2017,Moura:TCST:2017,Bartlett:TCST:2016,Lin:TCST:2015,Hu:AE:2020,Zhao:TCST:2017,RE:TIE:2014,Fang:CEP:2014}, the SoC does not align with the cell's actual level of energy in watt-hours (Wh)~\cite{Mamadou:ECS:2010}. As a case in point, the amount of energy the cell can deliver in the low SoC range is less than that in the high SoC range. By definition, the SoC also factors out the rate-capacity effect, which refers to the phenomenon of the deliverable charge or energy capacity becoming less (resp. more) under higher (resp. lower) discharging C-rates~\cite{Quade:B&S:2023}.

Compared to the SoC, the RDE comes as a more direct measure of the cell's remaining energy. RDE prediction has attracted growing interest, and the studies to date fall into two categories. The first category regards RDE as a synonym of SoC, naming it as state-of-energy (SoE). Firstly proposed in~\cite{Mamadou:ECS:2010}, SoE is defined as the percentage ratio of the cell's present energy capacity in Wh over the maximum energy capacity in Wh~\cite{Mamadou:ECS:2010}. The challenge in SoE estimation mainly lies in SoE modeling. One way is to relate the SoE with OCV~\cite{Wang:JPS:2016} and SoC~\cite{Zheng:AE:2016, Zhang:AE:2018, Ma:JES:2021} and then compute it based on these quantities. Another way is to build dynamic SoE models supplemented with the measurements of current, voltage, and temperature~\cite{Dong:Energy:2015}. Given these models, one can apply various state estimation methods, such as Kalman filtering~\cite{Zhang:JPS:2015,He:AE:2015} and particle filtering~\cite{Xiong:TIE:2018, Wang:AE:2020}, to estimate SoE. However, SoE estimation has not accounted for the effects of different discharging conditions, e.g., C-rates and temperature, on the actual energy availability~\cite{Dong:Energy:2015}.


The second category considers a literal definition of RDE---the integral of future discharging power (the product of current and voltage) over time. This approach runs the cell's dynamic model forward under a specific future current load until certain cut-off limits are reached and then performs numerical integration to determine the RDE. Some studies pursue the prediction of future current profiles by weighted moving average~\cite{Quinones:JPS:2018} and Markov modeling~\cite{Pola:ITR:2015,Niri:JES:2020}, and subsequently incorporate these predictions into the RDE calculation. However, the methods in this category generally focus on discharging at low to medium C-rates. As another limitation, the model forward simulation requires substantial amounts of computation, even though equivalent circuit models (ECMs) are commonly used.





Despite the many promising developments in RDE prediction, the state of the art is still limited. Existing works generally consider discharging at low to medium currents. In this case, the RDE is almost linear with the SoC~\cite{Quade:B&S:2023}, so the estimation of RDE bears little difference from SoC estimation. However, more interesting and useful is RDE prediction for discharging that spans low to high currents. This problem conveys more challenges while proving crucial for some emerging applications like electric aircraft~\cite{Yang:Joule:2021}, but the literature is void of studies to deal with it. We also note that some RDE prediction methods, e.g., those based on model-based forward simulation,  require heavy computation. Practical applications, however, demand methods that are computationally efficient and amenable to implementation. 

Motivated to overcome the above limitations, this paper, for the first time, explores the intricate task of RDE prediction over broad C-rate ranges and develops a solution framework based on machine learning (ML).  The contribution is threefold. 

\begin{itemize}

\item We propose a new RDE definition aligned with the considered problem. Different from its counterparts in the literature, the definition makes RDE dependent on C-rates and accounts for the effects of both the voltage limit and temperature limit. 

\item Following the definition, we develop an ML approach to make RDE prediction. The proposed ML architecture combines different learning modules, which are tasked to predict first the remaining time for discharging and then the remaining energy under specified C-rates and temperature and voltage limits. We provide the technical rationale to justify the design of each learning module. 

\item We propose to train the proposed ML approach on synthetic data, as otherwise the training would need a large number of discharging tests. To obtain high-fidelity synthetic data, we develop a separate hybrid physics+ML model capable of delivering accurate voltage and temperature prediction over broad current ranges. This hybrid model is an extension of our prior work in~\cite{Tu:AE:2023}, which is trained on experimental datasets.

\end{itemize}

The proposed RDE prediction approach is accurate, tractable for training, and computationally fast. We validate its effectiveness through extensive experiments on a nickel-cobalt-aluminum (NCA) cell and a lithium-iron-phosphate (LFP) cell.

Tangentially related with our work is using ML for other battery management tasks, such as SoC estimation, remaining useful life prediction, and fault detection. An interested reader is referred to~\cite{NOZARIJOUYBARI:JPS:2024,Aykol:JES:2021,GUO:JES:2022,Thelen:npj:2024} and the references therein. Our work in this paper, however, is distinct from the literature in two ways. First, we consider the new problem of RDE prediction when the current ranges from low to high C-rates, the complexity of which makes ML especially useful. Second, we present a unique, customized ML design for the considered problem.

The remainder of the paper is organized as follows. Section 2 proposes the definition of the C-rate-dependent RDE. Section 3 presents the hybrid physics-ML model to predict a LiB cell's voltage and temperature in discharging over broad current ranges. Section 4 presents an ML-based RDE prediction approach. Then, the experimental validation results are shown in Section 5. Section 6 provides our remarks about the results. Finally, Section 7 concludes the paper.

\section{Definition of C-rate-dependent RDE} \label{Sec: RDE definition}

This section proposes a new definition of RDE for LiBs. Different from other versions, this definition is designed to capture the variation of a cell's remaining available energy over different C-rates. Accurate and real-time prediction of the proposed RDE will be the aim of the subsequent sections.

\begin{figure}[t!]
    \centering
    \includegraphics[width = .48\textwidth,trim={10.5cm 3.7cm 10.4cm 4.3cm},clip]{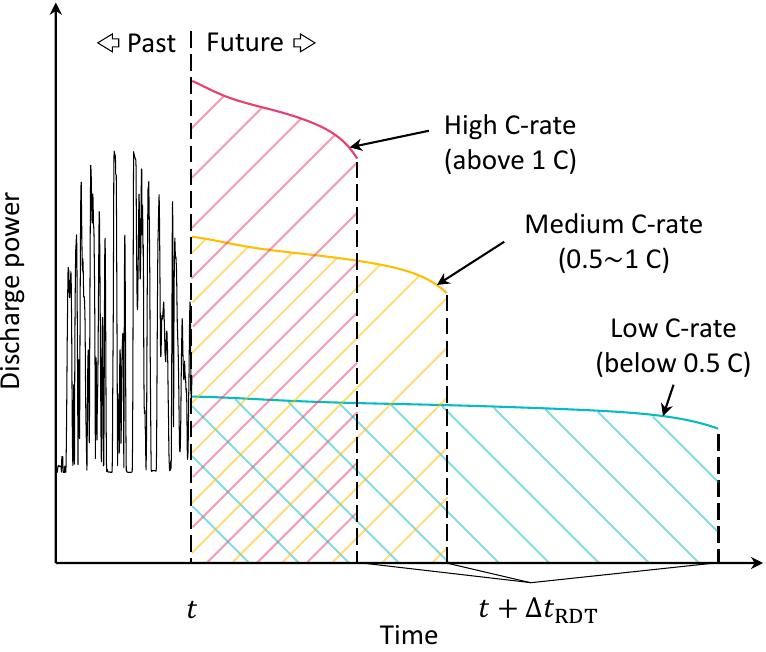}
    \caption{Illustration of the proposed C-rate-dependent RDE definition for LiBs. The shaded areas represent the RDE at different discharging C-rate levels. The present time is $t$ and the remaining discharge time is $\Delta t_{\mathrm{RDT}}$.}
    \label{Fig: RDE concept}
\end{figure}

\begin{figure*}[t!]
    \subfloat[]{
    \centering
    \includegraphics[width = .325\textwidth,trim={.3cm .6cm 1.34cm .7cm},clip]{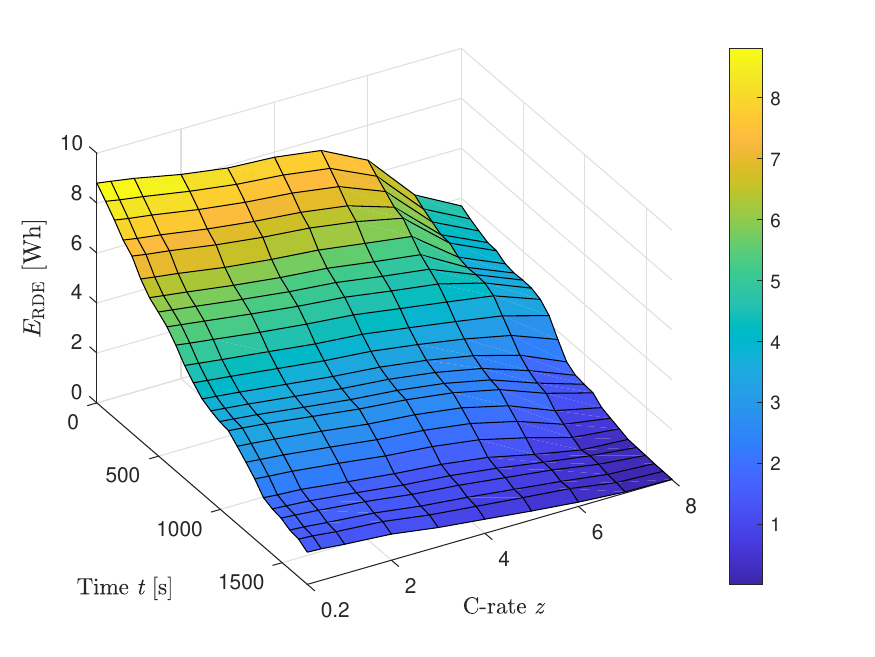}
    \label{Fig: US06-RDET-NCA}}
    \subfloat[]{
    \centering
    \includegraphics[width = .325\textwidth,trim={.3cm .6cm 1.34cm .7cm},clip]{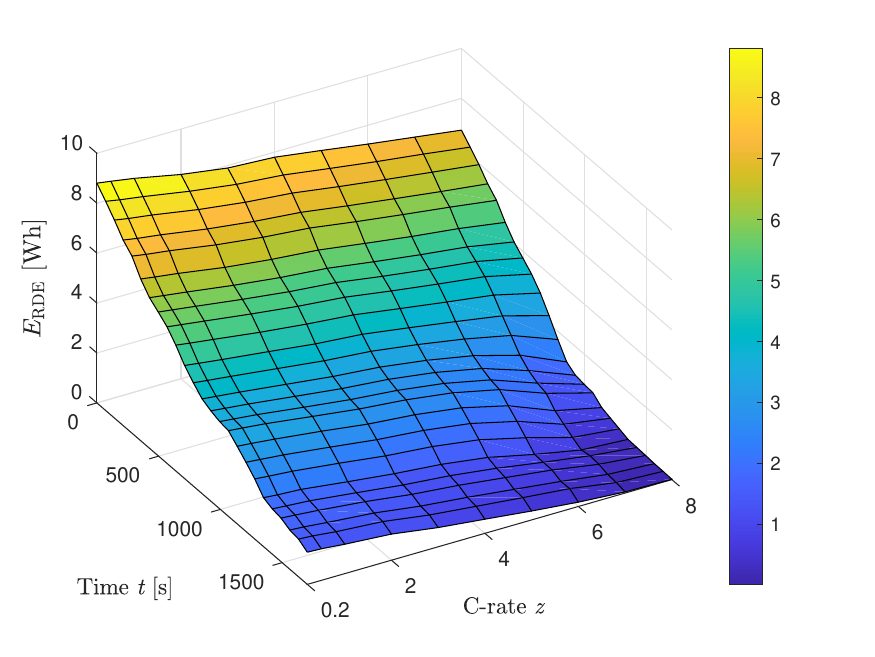}
    \label{Fig: US06-RDEV-NCA}}
    \subfloat[]{
    \centering
    \includegraphics[width = .325\textwidth,trim={.3cm .6cm 1.34cm .7cm},clip]{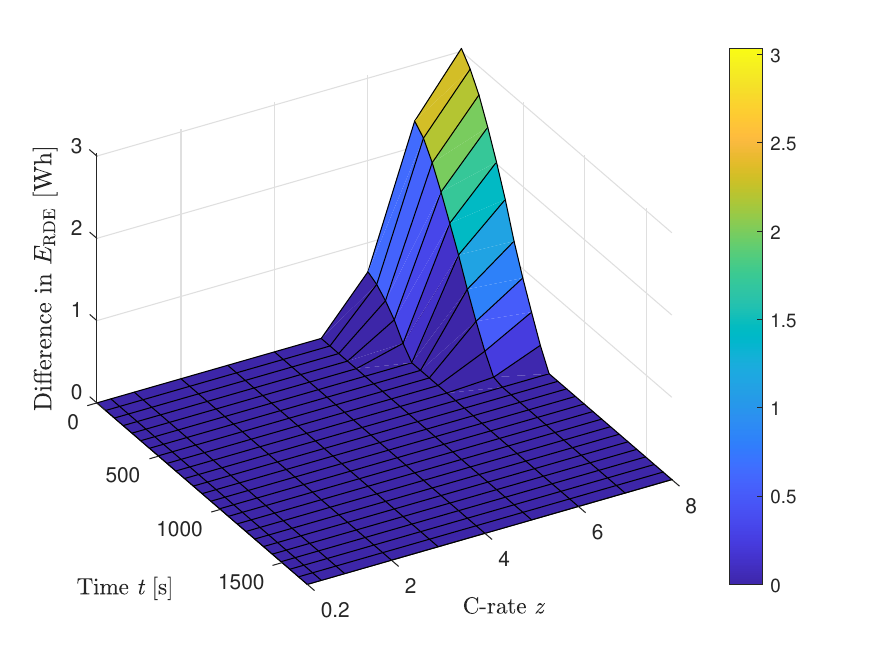}
    \label{Fig: US06-RDEdiff-NCA}}
    
    \caption{Variation of the RDE of an NCA cell across different C-rates under a modified US06 discharging profile, with $V_\mathrm{min} = 3\ \mathrm{V}$, $T_\mathrm{max} = 50 ^\circ\mathrm{C}$, and $T_\mathrm{amb} = 25 ^\circ\mathrm{C}$. (a) $E_\mathrm{RDE}$ subject to the limits of $V_{\mathrm{min}}$ and $T_{\mathrm{max}}$; (b) $E_\mathrm{RDE}$ subject to only the limit of $V_{\mathrm{min}}$; (c) difference of $E_\mathrm{RDE}$ between (a) and (b). The results show the C-rate dependence of a cell's RDE and demonstrate the effect of $T_{\mathrm{max}}$ limit for RDE prediction at high discharging C-rates.  When only considering $V_\mathrm{min}$ limit, the cell's temperature reaches a maximum of $67.5 ^\circ\mathrm{C}$ during the prediction of $E_\mathrm{RDE}$ at $z = 8\ \mathrm{C}$ and $t=0\ \mathrm{s}$.}
    \label{Fig: VT difference}
    
\end{figure*}

In general, the energy that a LiB cell will release over an upcoming time interval is the integration of its discharging power. This quantity, however, is neither constant nor fixed. Indeed, it is dependent on the discharging C-rate as a result of two main factors. First, the cell's output voltage will decline faster when the discharging current increases, leading to less energy to be released. This phenomenon is known as the rate-capacity effect. Second, the C-rate will greatly influence the cell's temperature. The discharging process and, consequently, the amount of energy that is to be discharged will be subject to the upper temperature limit for safety. RDE's C-rate dependence is too significant to be negligible at high C-rates, demanding a new definition of RDE to fit with discharging over broad current ranges. To proceed, we consider the RDE as the amount of the remaining energy if the cell is discharged at a constant C-rate from the present time until when either the voltage or the temperature limit is reached. With this notion, the RDE is defined as follows:
\begin{align}\label{Eqn:RDE}
    E_\mathrm{RDE}(z,t)=\int_{t}^{t + \Delta t_{\mathrm{RDT}}} zc_oV(\tau) d\tau,
\end{align}
where $t$ is the present time, $\Delta t_{\mathrm{RDT}}$ is the remaining discharging time (RDT), $z$ is the future discharge C-rate between $t$ and $t +\Delta t_{\mathrm{RDT}}$, $c_o$ is the current magnitude in ampere when $z = 1$ C, $V$ is the cell's voltage. Note that $t + \Delta t_{\mathrm{RDT}}$ is the time when the cell reaches the cut-off voltage $V_\mathrm{min}$ or the maximum allowed temperature $T_{\mathrm{max}}$, whichever earlier, if discharged at the constant current of $z$ C. Figure~\ref{Fig: RDE concept} illustrates this proposed RDE definition.

The RDE definition in~(\ref{Eqn:RDE}) is a departure away from alternative definitions in the literature,  e.g.,~\cite{Quade:B&S:2023}, which overlook the effects of the C-rates. Explicitly included in~(\ref{Eqn:RDE}), the C-rate $z$ impacts $E_\mathrm{RDE}$, due to its role in changing the evolution of $V(\tau)$ for $t < \tau \leq t+\Delta t_{\mathrm{RDT}}$ and in influencing $\Delta t_{\mathrm{RDT}}$. Figure~\ref{Fig: US06-RDET-NCA} illustrates the substantial variation of an NCA cell's $E_\mathrm{RDE}$ under different $z$ and subject to $V_\mathrm{min}$ and $T_\mathrm{max}$. It is also interesting to examine how much $T_\mathrm{max}$ affects $E_\mathrm{RDE}$, so Figure~\ref{Fig: US06-RDEV-NCA} shows $E_\mathrm{RDE}$ when $T_\mathrm{max}$ is disregarded. A significant difference from Figure~\ref{Fig: US06-RDET-NCA} emerges at high C-rates, which is depicted in Figure~\ref{Fig: US06-RDEdiff-NCA}. This observation underscores the considerable influence of high C-rates on the cell's thermal behavior, while emphasizing the importance of considering $T_\mathrm{max}$.

Given the proposed definition, predicting the RDE for a LiB cell involves multiple challenges. First, we need a model that is capable of accurately predicting the cell's voltage and temperature over broad current ranges. Second, even after the model is available, $\Delta t_\mathrm{RDT}$ can still be hard to determine as it depends on a mix of the cell's present state, future behavior at designated $z$, and pre-specified $V_\mathrm{min}$ and $T_\mathrm{max}$. Finally, online prediction of the RDE is desired, but how to achieve real-time computational efficiency is a question. In the sequel, we will develop a methodical approach to tackle these challenges by utilizing the capabilities of ML. An overview of our study is shown in Figure~\ref{Fig: Diagrams - Model and RDE}.

\begin{remark}
The dependence of a cell's capacity on the discharge C-rate is well-documented in the literature~\cite{DOERFFEL:JPS:2006}. One empirical method to capture this relationship is the Peukert equation. However, this equation lacks accuracy because it does not account for the cell's dynamics, nor does it incorporate $V_\mathrm{min}$ and $T_\mathrm{max}$ into the capacity estimation. Meanwhile, the Peukert equation addresses the charge capacity in Ah, but our RDE definition focuses on the energy capacity in Wh. This latter measure is more crucial for practical applications, yet also more challenging to determine due to the voltage dynamics.

 
\end{remark}

\section{Hybrid Physics-ML Modeling for LiBs}\label{Sec: LiB modeling}

\begin{figure}[t!]

    \subfloat[]{
    \centering
    \includegraphics[width = 0.5\textwidth,trim={4.5cm 7cm 6.5cm 3.8cm},clip]{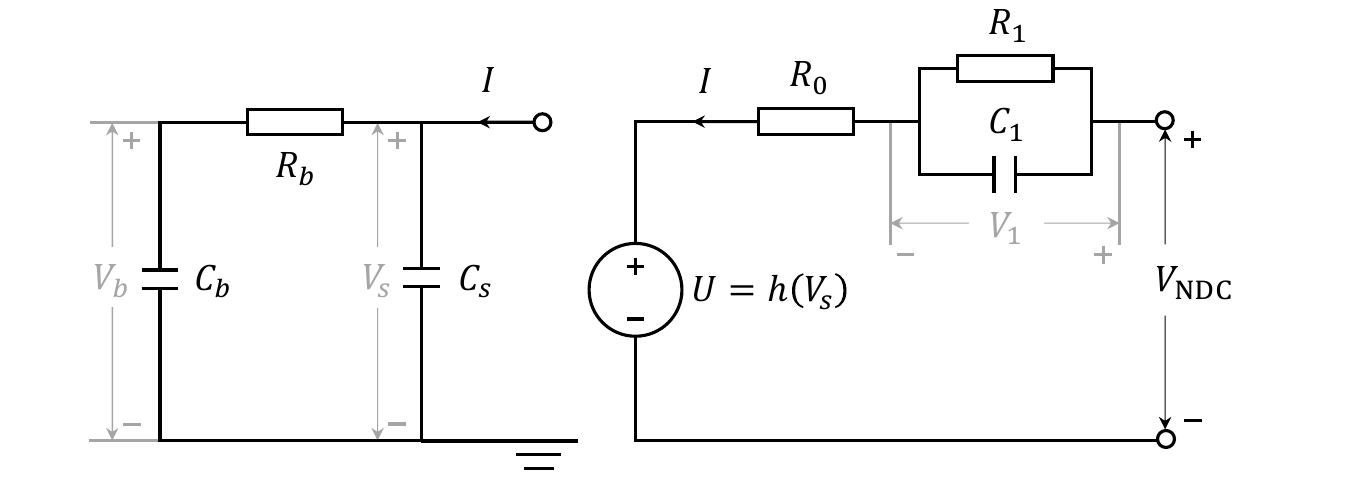}
    \label{Fig: NDC}
    }
    
    \subfloat[]{
    \centering
    \includegraphics[width = 0.5\textwidth,trim={3.5cm 7cm 9.5cm 7cm},clip]{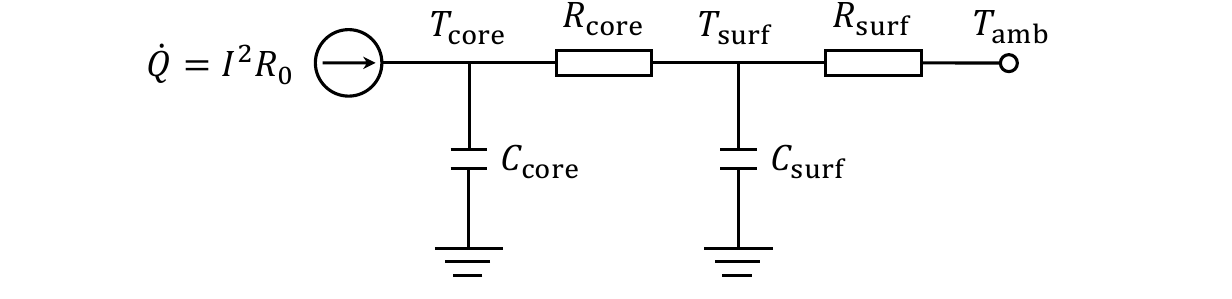}
    \label{Fig: Thermal}}
    
    \caption{ Diagrams of (a) the NDC model and (b) the lumped thermal model.}

\end{figure}

\begin{figure*}[t!]

    \subfloat[]{
    \centering
    \includegraphics[width = 0.99\textwidth,trim={.7cm 17.6cm .7cm .9cm},clip]{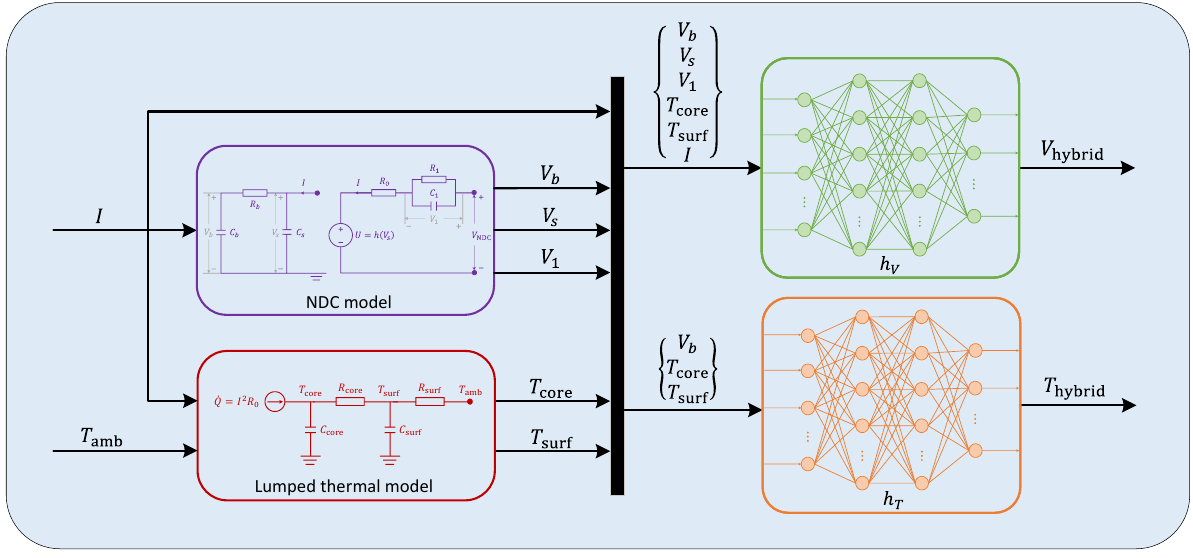}
    \label{Fig: Diagram - Model}}

    \subfloat[]{
    \centering
    \includegraphics[width = 0.99\textwidth,trim={.7cm 7.2cm .7cm 11.4cm},clip]{Diagram_v1.pdf}
    \label{Fig: Diagram - RDE}}
    
    \caption{(a) The NDCTNet model, which integrates physics with ML to predict the terminal voltage and temperature of a LiB cell, see Section~\ref{Sec: LiB modeling}; (b) the pipeline chart of the proposed RDE prediction approach, see Section~\ref{Sec: RDE prediction}.}
    \label{Fig: Diagrams - Model and RDE}

\end{figure*}

This section aims to develop a dynamic model to predict the terminal voltage and temperature of a LiB cell with fast computation and high accuracy when the cell discharging spans broad C-rate ranges. This model will be used to generate high-fidelity synthetic data to train the RDE prediction approach to be proposed in Section~\ref{Sec: RDE prediction}. We take the approach of integrating physics with ML, following our prior study~\cite{Tu:AE:2023}. For physics-based modeling, we use the nonlinear double capacitor (NDC) model~\cite{Tian:TCST:2021} for voltage prediction and a lumped thermal circuit model~\cite{Lin:JPS:2014} for temperature prediction.

We begin by introducing the NDC model and the lumped thermal model. The NDC model describes a LiB cell's electric behavior and, different from other ECMs, includes a mechanism to emulate the impact of lithium-ion diffusion on the voltage dynamics. As shown in Figure~\ref{Fig: NDC}, the model contains two coupled sub-circuits. The first (left) sub-circuit contains an R-C chain composed of $R_b$, $C_b$ and $C_s$. Conceptually, $C_b$ and $C_s$ represent the bulk inner region and surface region of the electrode, respectively; the charge migration between $C_b$ and $C_s$ mimics the lithium-ion diffusion within the electrode. The second (right) sub-circuit contains a voltage source $U$, a resistor $R_0$, and an R-C pair $R_1$-$C_1$. Here, $U=h(V_s)$ serves as an open-circuit voltage (OCV) source, $R_0$ corresponds to the ohmic resistance and solid electrolyte interface resistance, and $R_1$-$C_1$ accounts for the voltage transients due to charge transfer on the electrode/electrolyte interface. The governing equations of the NDC model are given by
\begin{subequations}
\begin{empheq}[left=\empheqlbrace]{align} \label{Eqn: NDC-Dynamic}
\begin{bmatrix}
\dot{V_b}(t) \\ \dot{V_s}(t) \\ \dot{V_1}(t)
\end{bmatrix}
&= A_{\mathrm{NDC}}
\begin{bmatrix}
V_b(t) \\ V_s(t) \\ V_1(t)
\end{bmatrix} 
+ B_{\mathrm{NDC}}I(t),\\ 
V_{\mathrm{NDC}}(t) &= h(V_s(t)) + V_1(t) + R_0 I(t),
\end{empheq}
\end{subequations}
where $V_b$, $V_s$ and $V_1$ are the voltages across $C_b$, $C_s$ and $C_1$, respectively, and $I$ is the input current ($I<0$ for discharge, $I>0$ for charge). Here, 
\begin{align*}
    A_{\mathrm{NDC}} =
    \begin{bmatrix}
    \frac{-1}{C_bR_b} & \frac{1}{C_bR_b} & 0 \\
    \frac{1}{C_sR_b} & \frac{-1}{C_sR_b} & 0 \\
    0 & 0 & \frac{-1}{R_1C_1}
    \end{bmatrix}, 
    B_{\mathrm{NDC}} =
    \begin{bmatrix}
    0\\
    \frac{1}{C_s}\\
    \frac{1}{C_1}
    \end{bmatrix}.
\end{align*}
The lumped thermal model, as shown in Figure~\ref{Fig: Thermal}, assumes a cylindrical shape of the cell and concentrates the spatially distributed temperature radially into two points at the cell's core and the surface, respectively. The model then is governed by
\begin{align} \label{Eqn: Thermal-Dynamic}
\begin{bmatrix}
\dot{T}_{\mathrm{core}}(t) \\ \dot{T}_{\mathrm{surf}}(t)
\end{bmatrix}
&= A_{\mathrm{therm}}
\begin{bmatrix}
T_{\mathrm{core}}(t) \\ T_{\mathrm{surf}}(t)
\end{bmatrix} 
+ B_{\mathrm{therm}}
\begin{bmatrix}
    I^2(t) \\
    T_{\mathrm{amb}}
\end{bmatrix},
\end{align}
where
\begin{align*}
    A_{\mathrm{therm}} &=
    \begin{bmatrix}
    \frac{-1}{R_{\mathrm{core}} C_{\mathrm{core}}} & \frac{1}{R_{\mathrm{core}} C_{\mathrm{core}}} \\
    \frac{1}{R_{\mathrm{core}} C_{\mathrm{surf}}} & \frac{-1}{R_{\mathrm{surf}} C_{\mathrm{surf}}}+\frac{-1}{R_{\mathrm{core}} C_{\mathrm{surf}}}
    \end{bmatrix}, \\
    B_{\mathrm{therm}} &=
    \begin{bmatrix}
    \frac{R_0}{C_{\mathrm{core}}} & 0\\
    0 & \frac{1}{R_{\mathrm{surf}} C_{\mathrm{surf}}}
    \end{bmatrix},
\end{align*}
$T_{\mathrm{core/surf}}$ is the temperature at the core/surface of the cell, $T_{\mathrm{amb}}$ is the ambient temperature, and $C_{\mathrm{core/surf}}$ is the heat capacity at the cell's core/surface. Further, $R_\mathrm{core}$ is the thermal resistance due to the conduction between the core and surface, and $R_\mathrm{surf}$ is the thermal resistance due to the convection between the cell's surface and the environment. This model only considers internal heat generation at the core that results from Joule heating.

The above two ECMs have decent accuracy at low to medium C-rates but lack accuracy for high C-rates. To overcome the gap, we couple them with ML models. ML has strong data-driven representation capabilities, and we have shown in our prior study~\cite{Tu:AE:2023} that they can predict the behaviors of a LiB cell with high accuracy if integrated with physics-based models. Based on the insights in~\cite{Tu:AE:2023}, we propose the NDCTNet model in Figure~\ref{Fig: Diagram - Model}, which characteristically connects the above two ECMs with two separate feedforward neural networks (FNNs) in series. The first FNN is responsible for the terminal voltage prediction. It takes $V_b$, $V_s$, $V_1$, $T_\mathrm{core}$, $T_\mathrm{surf}$ and $I$ as the inputs. The second FNN is tasked to predict the surface temperature, and its input includes $V_b$, $T_\mathrm{core}$ and $T_\mathrm{surf}$.  Here, the FNNs are made to use physical state information for prediction. This allows to use simpler FNNs, reduce data dependence in training, and achieve more accurate prediction.  It should be noted that the inputs to the FNNs must be selected so that the FNNs can capture physically justifiable mappings~\cite{Tu:AE:2023}.  The proposed choice is the best one that we could find after many trials, though there may exist alternatives.

In summary, the NDCTNet model can be written in the following mathematical form:
\begin{subequations} \label{Eqn: hybrid model}
\begin{empheq}[left=\empheqlbrace]{align} \label{Eqn: hybrid dynamic model}
      &\dot{x} = f_{\mathrm{phy}}(x,I,T_{\mathrm{amb}}), \\
    &V_{\mathrm{hybrid}} = h_V(x,I), \\
    &T_{\mathrm{hybrid}} = h_T(x),
\end{empheq}
\end{subequations}
where $x = [V_b\ V_s\ V_1\ T_\mathrm{core}\ T_\mathrm{surf}]^\top$, (\ref{Eqn: hybrid dynamic model}) results from (\ref{Eqn: NDC-Dynamic}) and (\ref{Eqn: Thermal-Dynamic}) and is a linear ordinary differential equation, and $h_V$ and $h_T$ represent the two FNNs.

The identification of the NDCTNet model follows a two-step procedure. First, we extract the NDC model and the thermal circuit model from experimental datasets collected at low- to medium-C-rate discharging. The data include current, voltage, and surface temperature, and the parameter estimation is achieved by fitting the models with the data in a least-squares manner~\cite{Tian:TCST:2021}. Next, we perform additional discharging tests across low to high C-rates and run the NDC model and the lumped thermal model under the same current profiles. We then train the FNNs, $h_{V}$ and $h_{T}$, using the new datasets. The reader is referred to~\cite{Tu:AE:2023} for more information.

\begin{remark}
    Various physics-based battery models have flourished in the literature, including electrochemical models~\cite{Krewer:JES:2018,Chaturvedi:CSM:2010} and ECMs~\cite{Hu:JPS:2012}.  These models provide physical interpretability, but struggle with either computational efficiency or predictive accuracy. Recently, ML  has emerged as a powerful tool for data-driven prediction of battery behaviors.  While computationally fast and accurate they are, ML models often lack physical consistency while requiring large amounts of training data. However, our previous study~\cite{Tu:AE:2023} shows that integrating physics-based modeling with ML can enable high accuracy, low computation, and reduced dependence on training data. The NDCTNet model extends the work in~\cite{Tu:AE:2023} by incorporating temperature prediction in addition to voltage prediction.
\end{remark}

\section{Real-Time RDE Prediction via NNs}\label{Sec: RDE prediction}

In this section, we focus on the problem of real-time RDE prediction for LiBs. Note that, given the LiB model developed in Section~\ref{Sec: LiB modeling}, a straightforward approach to the RDE prediction is to forward run the NDCTNet model from the present state until $V_\mathrm{min}$ or $T_\mathrm{max}$ is reached, as is reported in some existing studies, e.g.,~\cite{Niri:JES:2020}. However, this approach requires high computational costs. This is because not only the model execution may take a long time, as the actual discharging can be up to hours, but also the model must be solved for different C-rates from low to high. Therefore, we propose an ML-based approach to enable real-time RDE prediction. This new approach will eliminate the need for repetitive forward model simulation as ML models can give predictions efficiently after being trained. Also, we can harness the capacity of ML to achieve high RDE prediction accuracy given abundant data.

Based on the definition of $E_\mathrm{RDE}$ in~(\ref{Eqn:RDE}), we aim to train an FNN which will output $E_\mathrm{RDE}$ based on the present state $x(t)$, $z$ and $\Delta t_\mathrm{RDT}$. Here, a key problem is to determine $\Delta t_\mathrm{RDT}$. As the RDT, $\Delta t_\mathrm{RDT}$ will depend on whichever of $V_\mathrm{min}$ and $T_\mathrm{max}$ comes first in discharging, implying
\begin{align} \label{Eqn: RDTVminTmax}
\Delta t_\mathrm{RDT} = \min \left\{\Delta t_\mathrm{RDT}^{V_\mathrm{min}}, \Delta t_\mathrm{RDT}^{T_\mathrm{max}}   \right\},
\end{align}
where $\Delta t_\mathrm{RDT}^{V_\mathrm{min}}$ (resp. $\Delta t_\mathrm{RDT}^{T_\mathrm{max}}$) is the time duration that elapses before the cell reaches $V_\mathrm{min}$ (resp. $T_\mathrm{max}$). It should be noted that both $\Delta t_\mathrm{RDT}^{V_\mathrm{min}}$ and $\Delta t_\mathrm{RDT}^{T_\mathrm{max}}$ are $z$-dependent. We find out that there exists a mapping from $x(t)$, $z$ and $T_\mathrm{amb}$ to $\Delta t_\mathrm{RDT}^{V_\mathrm{min}}$, which can be captured by an FNN. However, $\Delta t_\mathrm{RDT}^{T_\mathrm{max}}$ is much more difficult to compute since the cell constantly has heat exchange with the ambient environment, and the ambient temperature, $T_\mathrm{amb}$, can be arbitrary. As it is intractable to train an FNN to predict $\Delta t_\mathrm{RDT}^{T_\mathrm{max}}$, we use the bisection method instead to find $\Delta t_\mathrm{RDT}^{T_\mathrm{max}}$ if the cell will exceed $T_\mathrm{max}$ before the elapse of $\Delta t_\mathrm{RDT}^{V_\mathrm{min}}$. To sum up, the proposed RDE prediction approach includes the following three steps:

\begin{itemize}

\item leverage an FNN to compute $\Delta t_\mathrm{RDT}^{V_\mathrm{min}}$ based on $x(t)$, $z$ and $T_\mathrm{amb}$;

\item let  $\Delta t_\mathrm{RDT}$ = $\Delta t_\mathrm{RDT}^{V_\mathrm{min}}$ if  $\Delta t_\mathrm{RDT}^{T_\mathrm{max}}>\Delta t_\mathrm{RDT}^{V_\mathrm{min}}$; otherwise, use the bisection search to find out $\Delta t_\mathrm{RDT}^{T_\mathrm{max}}$, and let $\Delta t_\mathrm{RDT}$ = $\Delta t_\mathrm{RDT}^{T_\mathrm{max}}$;

\item  given $\Delta t_\mathrm{RDT}$, use an FNN to predict $E_\mathrm{RDE}$ based on $x(t)$, $z$ and $T_\mathrm{amb}$.

\end{itemize}
Below we present the details of each step.

\subsection{Computation of $\Delta t_\mathrm{RDT}^{V_\mathrm{min}}$} \label{Sec: RDE - Step 1}

The first step aims to use an FNN to predict $\Delta t_\mathrm{RDT}^{V_\mathrm{min}}$  based on the triplet ($x(t)$, $z$, $T_\mathrm{amb}$). However, one may wonder whether FNNs are able to perform the task. It is known that FNNs are suitable for approximating continuous mappings due to the universal approximation theorem~\cite{Hornik:NN:1989}, so the question is whether there exists such a mapping from ($x(t)$, $z$, $T_\mathrm{amb}$) to a unique $\Delta t_\mathrm{RDT}^{V_\mathrm{min}}$. To find the answer, we conduct the following analysis. 

We begin with considering~(\ref{Eqn: hybrid model}). Note that~(\ref{Eqn: hybrid dynamic model}) is a linear time-invariant system and thus has a closed-form solution. Letting  $x(t)$ be the initial condition, we express the solution $x(t+\Delta t)$  for an arbitrary time interval $\Delta t>0$ under given $z$ and $T_\mathrm{amb}$ as
\begin{align*}
x(t+\Delta t) = \phi ( x(t), zc_o, T_\mathrm{amb}; \Delta t),
\end{align*}
where the exact formula for $\phi$ is available, see Appendix. Then, 
\begin{align}\label{Eqn: Vhybridhvphi}
V_\mathrm{hybrid}(t+\Delta t)  = h_V \circ \phi(x(t), zc_o,T_\mathrm{amb}; \Delta t),
\end{align}
where $\circ$ denotes the operation of function composition. We further define $h_V^\phi = h_V \circ \phi$ and absorb $c_o$ into it for notational simplicity. Then, $\Delta t_\mathrm{RDT}^{V_\mathrm{min}}$ must satisfy
\begin{align} \label{Eqn: RDTVmin}
h_V^\phi(x(t), z, T_\mathrm{amb}; \Delta t_\mathrm{RDT}^{V_\mathrm{min}}) = V_\mathrm{min}.
\end{align}
The following observations arise out of the equality in~(\ref{Eqn: RDTVmin}).  First,  by selecting a suitable activation function for the NNs, $h_V^\phi$ is continuously differentiable because both $h_V$ and $\phi$ are continuously differentiable. Second, for any physically possible ($x(t)$, $z$, $T_\mathrm{amb}$), there always exists a unique $\Delta t_\mathrm{RDT}^{V_\mathrm{min}}$ in discharging to make (\ref{Eqn: RDTVmin}) hold true. Third, from a physical perspective, it is arguable that an open neighborhood of ($x(t)$, $z$, $T_\mathrm{amb}$) would correspond to an open neighborhood of $\Delta t_\mathrm{RDT}^{V_\mathrm{min}}$. 
Finally, when the cell is continuously discharged starting from $x(t)$ and under given $z$ and $T_\mathrm{amb}$, its terminal voltage will decline monotonically when the discharging time approaches $\Delta t_\mathrm{RDT}^{V_\mathrm{min}}$. This implies 
\begin{align*}
\left. \frac{\partial h_V^\phi }{\partial \Delta t} \right|_{\Delta t = \Delta t_\mathrm{RDT}^{V_\mathrm{min}}}<0. 
\end{align*}
Then, by the global implicit function theorem in~\cite{Sanberg:TCS:1981}, there is a unique and continuous mapping from ($x(t)$, $z$, $T_\mathrm{amb}$) to $\Delta t_\mathrm{RDT}^{V_\mathrm{min}}$. With the existence of such a mapping, an FNN can be used to predict $\Delta t_\mathrm{RDT}^{V_\mathrm{min}}$ based on ($x(t)$, $z$, $T_\mathrm{amb}$), which is denoted as
\begin{align*}
    \Delta t_\mathrm{RDT}^{V_\mathrm{min}} = \mathrm{FNN}_\mathrm{RDT}^{V_\mathrm{min}}(x(t), z, T_\mathrm{amb}).
\end{align*}

\subsection{Computation of $\Delta t_\mathrm{RDT}^{T_\mathrm{max}}$}\label{Sec: RDT with temperature}

\begin{figure}[t!]
\centering
\resizebox{0.98\columnwidth}{!}{
\tikzstyle{start} = [rectangle, rounded corners, 
minimum width=3.5cm, 
minimum height=1.2cm,
text centered, 
draw=black] 

\tikzstyle{stop} = [rectangle, rounded corners, 
minimum width=3cm, 
minimum height=1.2cm,
text centered, 
draw=black] 

\tikzstyle{process1} = [rectangle, 
minimum width=3cm, 
minimum height=1.2cm, 
text centered, 
text width=5.0cm, 
draw=black]

\tikzstyle{process1.1} = [rectangle, 
minimum width=5.6cm, 
minimum height=1.2cm, 
text centered, 
text width=6.8cm, 
draw=black]

\tikzstyle{process2} = [rectangle, 
minimum width=3cm, 
minimum height=1cm, 
text centered, 
text width=4cm, 
draw=black] 

\tikzstyle{process3} = [rectangle, 
minimum width=1cm, 
minimum height=1cm, 
text centered, 
text width=2cm, 
draw=black] 

\tikzstyle{decision} = [diamond, 
aspect=1.4,
minimum width=0 cm, 
minimum height=0 cm, 
text centered, 
draw=black] 

\tikzstyle{arrow} = [thick,->,>=stealth]

\begin{tikzpicture}[node distance=2cm]

\node (start) [start] {\makecell[c]{Input \\ $x(t)$, $z$, $T_\mathrm{amb}$, $T_\mathrm{hybrid}$}};
\node (pro0) [process1, below of=start,yshift=+0.15cm] {Pick $m$ checkpoints $\delta_i$ for $i=1,\ldots,m$ between $\left[0, \Delta t_\mathrm{RDT}^{V_\mathrm{min}}\right]$};
\node (dec0) [decision, below of=pro0, yshift=-1.1cm] {\makecell[c]{ $T_\mathrm{hybrid}(t+\delta_i)$ $<T_\mathrm{max}$\\ for all $i =1,\ldots,m$ ?}};
\node (pro1) [process1.1, below of=dec0, yshift=-1.1cm] {Set $\Delta \underline{t} = \delta_{i^*-1}$ and $\Delta \bar {t} = \delta_{i^*}$, with $i^* = \min \left\{ i \; | \; T_\mathrm{hybrid}(t+\delta_i)>T_\mathrm{max}, i=1,\ldots,m \right\}$};
\node (pro2) [process2, below of=pro1, yshift=+0.2cm] {Set $\tau = (\Delta \underline{t} + \Delta \bar{t}) /2$};
\node (dec2) [decision, below of=pro2, yshift=-1.1cm] {\makecell[c]{$|T_\mathrm{hybrid}(t+\tau) - T_\mathrm{max}| $$ < \epsilon$?}};
\node (dec1) [decision, below of=dec2, yshift=-2.2cm] {\makecell[c]{$T_\mathrm{hybrid}(t+\tau)$ $<T_\mathrm{max}$?}};
\node (pro3b) [process3, below of=dec1, yshift=-0.8cm] {$\Delta \bar{t}=\tau$};
\node (pro3a) [process3, right of =dec1, xshift=2.2cm] {$\Delta \underline{t}=\tau$};
\node (out1) [stop, right of=dec0, xshift=2.85cm] {\makecell[c]{Output \\ $\Delta t_\mathrm{RDT} = \Delta t_\mathrm{RDT}^{V_\mathrm{min}}$}};
\coordinate[below of = pro3b, yshift=+0.9cm] (aux2);  
\node (out2) [stop, right of=dec2, xshift=2.9cm] {\makecell[c]{Output\\ $\Delta t_\mathrm{RDT}^{T_\mathrm{max}} = \tau$}};

\draw [arrow] (start) -- (pro0);
\draw [arrow] (pro0) -- (dec0);
\draw[arrow] (dec0) -- node[pos=0.4,right] {No} (pro1);
\draw[arrow] (pro1) -- (pro2);
\draw[arrow] (pro2) -- (dec2);
\draw[arrow] (dec2) -- node[pos=0.4,right] {No} (dec1);
\draw[arrow] (dec1) -- node[pos=0.4,right] {No} (pro3b);
\draw[arrow] (dec1) -- node[pos=0.4,above] {Yes} (pro3a);
\draw[arrow] (pro3b) -- (aux2) -| (-4,-20) |- (pro2);
\draw[arrow] (pro3a) |- (aux2);
\draw[arrow] (dec2) -- node[pos=0.4,above] {Yes} (out2);
\draw[arrow] (dec0) -- node[pos=0.4,above] {Yes} (out1);

\end{tikzpicture}}
\caption{Bisection search for $\Delta t_\mathrm{RDT}^{T_\mathrm{max}}$.}
\label{Fig: RDT prediction}
\end{figure}

In predicting the RDE, we must take the cell's future temperature into account to ensure that the prediction lies within the thermal safety bounds. This is particularly vital when the cell runs at high C-rates or under high ambient temperature. We have defined $\Delta t_\mathrm{RDT}^{T_\mathrm{max}}$ earlier to capture the RDT due to $T_\mathrm{max}$. Note that $\Delta t_\mathrm{RDT}^{T_\mathrm{max}}$ needs to be computed for the RDE prediction only when it is less than $\Delta t_\mathrm{RDT}^{V_\mathrm{min}}$, as implied by~(\ref{Eqn: RDTVminTmax}). Therefore, it is plausible to expediently determine whether $\Delta t_\mathrm{RDT}^{T_\mathrm{max}}<\Delta t_\mathrm{RDT}^{V_\mathrm{min}}$, and in case of not, a fast search for $\Delta t_\mathrm{RDT}^{T_\mathrm{max}}$ will ensue next. 

To check $\Delta t_\mathrm{RDT}^{T_\mathrm{max}}<\Delta t_\mathrm{RDT}^{V_\mathrm{min}}$, we pick $m$ checkpoints,  $\delta_i$ for $i=1,\ldots,m$, which spread evenly within the interval $\left[0, \Delta t_\mathrm{RDT}^{V_\mathrm{min}}\right]$.   One may choose an appropriate  $m$, depending on the characteristics of the cell's thermal dynamics and the immediate ambient conditions. Then, $T_\mathrm{hybrid}(t+\delta_i)$ is computed for each $\delta_i$ via
\begin{align*}
T_\mathrm{hybrid}(t+\delta_i) = \underbrace{h_T \circ \phi}_{h_T^\phi}  (x(t), z, T_\mathrm{amb}; \delta_i).
\end{align*}
The computation would be efficient, thanks to the closed-form formula of $\phi$. We can skip the calculation of $\Delta t_\mathrm{RDT}^{T_\mathrm{max}}$  if $T_\mathrm{hybrid}(t+\delta_i)<T_\mathrm{max}$ for all $i=1,\ldots,m$. Otherwise, we must determine $\Delta t_\mathrm{RDT}^{T_\mathrm{max}}$, and this is equivalent to finding the root of
\begin{align} \label{T_max-Equality}
T_\mathrm{hybrid}(t+\delta t) = h_T^\phi(x(t), z, T_\mathrm{amb}; \delta t) = T_\mathrm{max}, 
\end{align}
for which $\delta t$ is the unknown variable. To achieve this, we apply the bisection method, a popular and efficient root-finding algorithm.

The bisection search is conducted within the interval $\left[\Delta \underline{t}, \Delta \bar {t}\right]$. Initially, $\Delta \bar{t} = \delta_{i^*}$ with $i^*=\min \left\{ i \; | \; T_\mathrm{hybrid}(t+\delta_i)>T_\mathrm{max}, i=1,\ldots,m \right\}$, and $\Delta \underline{t} = \delta_{i^*-1}$. Assume that there is only one root for~(\ref{T_max-Equality}) within the initial interval. Subsequently, we bisect the interval, evaluate $T_\mathrm{hybrid}$ at the midpoint, identify the next subinterval to search, and repeat the search procedure. The steps are as follows.
\begin{enumerate}[ {[}Step 1{]} ]

\item calculate $\tau = (\Delta \underline{t} + \Delta \bar{t}) /2$, and evaluate $T_\mathrm{hybrid}(t+\tau)$;

\item let $\Delta t_\mathrm{RDT}^{T_\mathrm{max}} = \tau$ and stop if $|T_\mathrm{hybrid}(t+\tau) - T_\mathrm{max}| < \epsilon$, where $\epsilon$ is the tolerance; otherwise, go to Step 3. 

\item let $\Delta \underline{t} = \tau$ if $T_\mathrm{hybrid}(t+\tau)<T_\mathrm{max}$, and let $\Delta \bar{t} = \tau$ if otherwise; go to Step 1. 

\end{enumerate}
Figure~\ref{Fig: RDT prediction} shows a flowchart of the procedure. The bisection search can find out $\Delta t_\mathrm{RDT}^{T_\mathrm{max}}$ fast here. This is because an explicit formula is available for the computation of $T_\mathrm{hybrid}(t+\tau)$, and the bisection method converges fast by itself when a unique root exists within the search interval.

\subsection{Prediction of $E_\mathrm{RDE}$} \label{Sec: RDE - Step 3}

After identifying $\Delta t_\mathrm{RDT}$, we are now ready to enable the prediction of $E_\mathrm{RDE}$. It is beneficial here to consider a more general task --- predicting $E(z, t, \Delta t)$, which is the discharging energy released by the cell within the upcoming time interval $(t, t+\Delta t]$ under a constant C-rate $z$:
\begin{align*} 
E(z, t, \Delta t) = \int_{t}^{t + \Delta t} zc_oV(\tau) d\tau.
\end{align*}
When $E(z, t, \Delta t)$ is available, we can easily compute $E_\mathrm{RDE}$ by inserting $\Delta t_\mathrm{RDT}$ into it
\begin{align} \label{Eqn: EtoRDE}
E_\mathrm{RDE}(z,t) = E(z,t,\Delta t_\mathrm{RDT}).
\end{align}
Given~(\ref{Eqn: Vhybridhvphi}), we have
\begin{align}
    E(z, t, \Delta t)
    =\int_{0}^{\Delta t} zc_o h_V^\phi(x(t), z, T_{\mathrm{amb}}; \tau) d\tau.
\end{align}
This relation indicates that there exists a continuous mapping from ($x(t)$, $z$, $T_\mathrm{amb}$, $\Delta t$) to $E$. While there is no analytical form of the mapping, we can use an FNN to approximate it. The FNN is denoted as
\begin{align*}
E(z, t, \Delta t) = \mathrm{FNN}_{E}(x(t), z, T_\mathrm{amb}, \Delta t).
\end{align*}
After the FNN is trained, we can compute $E_\mathrm{RDE}$ by~(\ref{Eqn: EtoRDE}).

Figure~\ref{Fig: Diagram - RDE} presents the pipeline of the proposed RDE prediction approach. To sum up, it is challenging to predict the RDE when the cell shows complex voltage and thermal behaviors under discharging across low to high currents. To overcome this problem, we turn our attention to utilizing the power of ML. We exploit FNNs to capture the sophisticated relationship between the RDE and the many factors that affect it. The approach presents several benefits. First, it is amenable to training. This is because we can train the FNNs on synthetic data generated by running the NDCTNet model in Section~\ref{Sec: LiB modeling}. This greatly reduces reliance on data collection from experiments, as now only the NDCTNet training requires experimental data. Second, the approach can attain high accuracy when the NDCTNet model is accurate enough to generate good-quality training data in large amounts. Finally, the prediction is computationally fast and efficient after the training, as we only need to run the FNNs and, if there is a need, the bisection search procedure.

The proposed ML approach is distinct from existing RDE prediction methods in multiple ways. First, it tackles the challenge of RDE prediction under discharging over broad current ranges, a topic that has not been studied before. Second, to deal with this complex problem, the approach presents a customized ML architecture, which includes and combines different learning modules into a holistic framework. Finally, the design of each module is justified in rationale to ensure its technical soundness, which has been rarely done in previous studies on RDE prediction.

\section{Experimental Validation} \label{Sec: Exp validation}

\begin{figure}[t!]
    \centering
    \includegraphics[width = .38\textwidth,trim={4cm 0cm 4cm 0cm},clip]{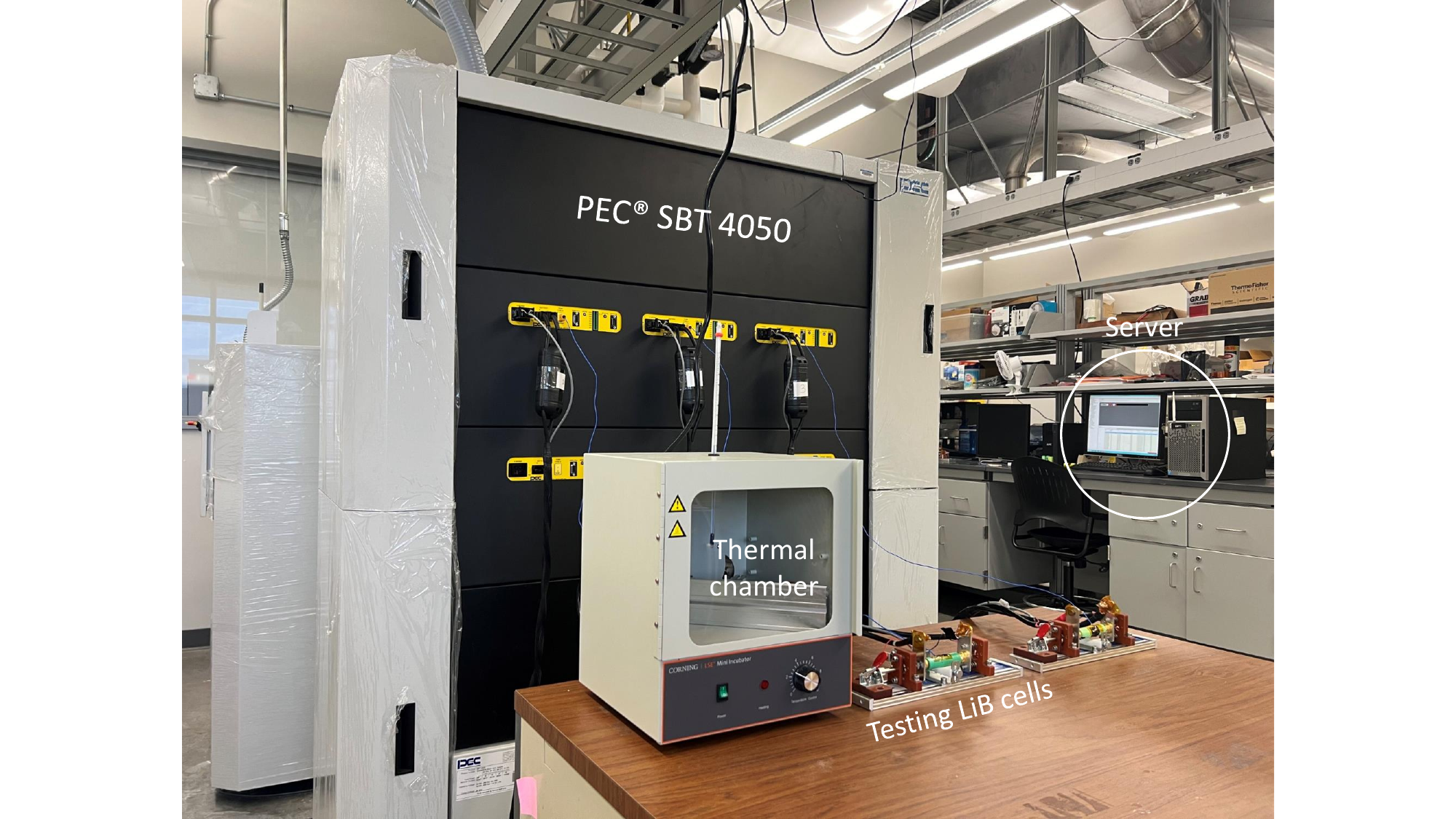}
    \caption{The PEC$\textsuperscript{\textregistered}$ SBT4050 battery tester.}
    \label{Fig: PEC tester}
\end{figure}

\begin{table}[t!]\centering
 \resizebox{.95\columnwidth}{!}{
 \begin{tabular}{ m{5.45cm}  m{1.05cm}  m{1.05cm} }
\toprule
 & NCA & LFP \\
\midrule
Nominal capacity (Ah) & 2.5 & 2.5 \\
Allowed voltage range (V) & 2.5$\sim$4.2 & 2.0$\sim$3.6 \\
Max. continuous discharge current (A) & 20 & 50 \\
Operating surface temperature ($^{\circ}\mathrm{C}$) & -20$\sim$75 & -30$\sim$55 \\
\bottomrule
\end{tabular}
}
\caption{Specifications of the NCA and LFP cells.}
\label{Table: LiB cells}
\end{table}

\begin{table}[t!]\centering

 \resizebox{.95\columnwidth}{!}{
 \begin{tabular}{ m{.9cm} p{3.6cm} m{1.4cm} m{1.4cm} }
\toprule
LiB type & \makecell{Current \\ profile} & RMSE $V_\mathrm{hybrid}$ & RMSE $T_\mathrm{hybrid}$ \\
\midrule
\multirow{2}{2cm}{NCA} &\makecell{ Modified US06 (0$\sim$8 C)} & 12.25 mV & 0.27 $^\circ\mathrm{C}$ \\
 & \makecell{ Modified SC04 (0$\sim$8 C)} & 13.36 mV & 0.56 $^\circ\mathrm{C}$ \\
 \midrule
\multirow{2}{2cm}{LFP} & \makecell{ Modified US06 (0$\sim$15 C)} & 13.01 mV & 0.28 $^\circ\mathrm{C}$ \\
 & \makecell{ Modified SC04 (0$\sim$15 C)} & 11.11 mV & 0.46 $^\circ\mathrm{C}$ \\
\bottomrule
\end{tabular}
}
\caption{Validation results for the two NDCTNet models constructed for the NCA and LFP cells.}
\label{Table: Model Validation}
\end{table}

In this study, the experimental validation is performed on two cylindrical LiB cells: a Samsung INR18650-25R cell with an NCA cathode and a graphite anode, and an A123 ANR26650-M1B cell with an LFP cathode and a graphite anode. Table~\ref{Table: LiB cells} lists their specifications. A PEC$\textsuperscript{\textregistered}$ SBT4050 battery tester shown in Figure~\ref{Fig: PEC tester} was used to charge/discharge the cells and collect experimental data. A thermocouple was attached to each cell's surface to measure its temperature. During the experiments, the cells were placed in a thermal chamber with the ambient temperature set to $T_{\mathrm{amb}}=25$$^{\circ}\mathrm{C}$. A host computer with a 2.2 GHz Intel$\textsuperscript{\textregistered}$ i7-8750H CPU and 16.0 GB RAM was used to process the collected data and implement the proposed RDE prediction approach.

The validation consists of two stages. In the first stage, we identify and validate the NDCTNet model constructed in Section~\ref{Sec: LiB modeling} using experimental data. In the second stage, we run the NDCTNet model in simulations to generate the synthetic data and train the RDE prediction approach. After the training, we implement the approach and compare its RDE prediction results with the experimental truth.

\subsection{Training and Validation of the NDCTNet Model} \label{Sec: Exp - NDCTNet}

\begin{figure*}[t!]

\centering
    \subfloat[]{
    \centering
    \includegraphics[width = .48\textwidth,trim={1cm .7cm 1cm .9cm},clip]{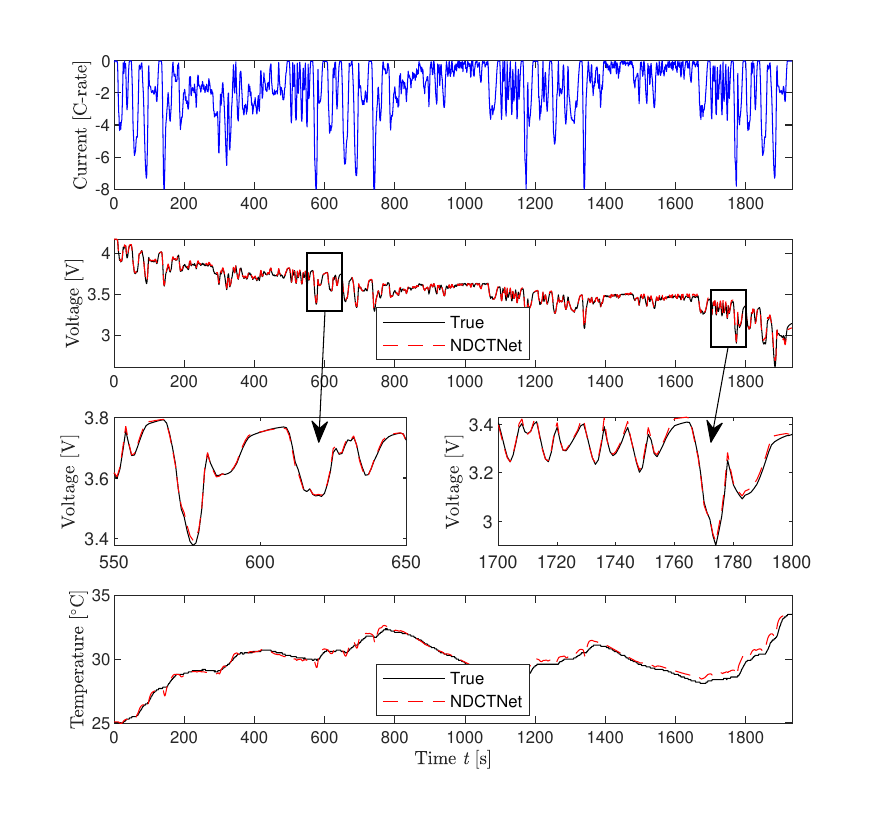}}
    \subfloat[]{
    \centering
    \includegraphics[width = .48\textwidth,trim={1cm .7cm 1cm .9cm},clip]{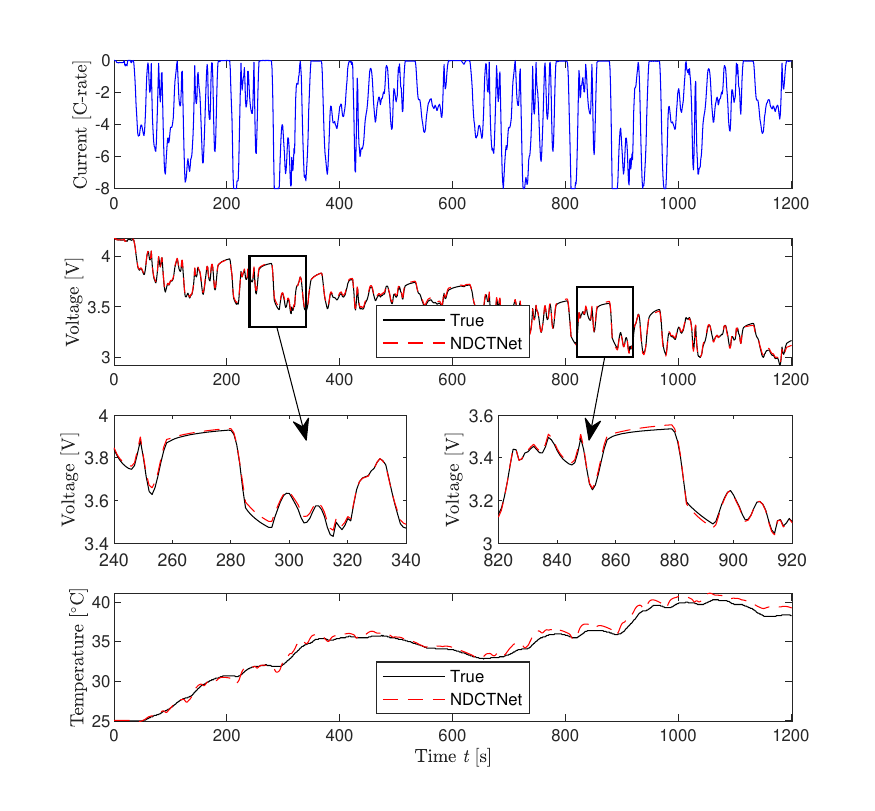}}
    \caption{Validation of the NDCTNet model for the NCA cell based on (a) the modified US06 (0$\sim$8 C) and (b) the modified SC04 (0$\sim$8 C) profiles.}
    \label{Fig: US06-Model-NCA}
    
\end{figure*}

\begin{figure*}[t!]

\centering
    \subfloat[]{
    \centering
    \includegraphics[width = .48\textwidth,trim={1cm .7cm 1cm .9cm},clip]{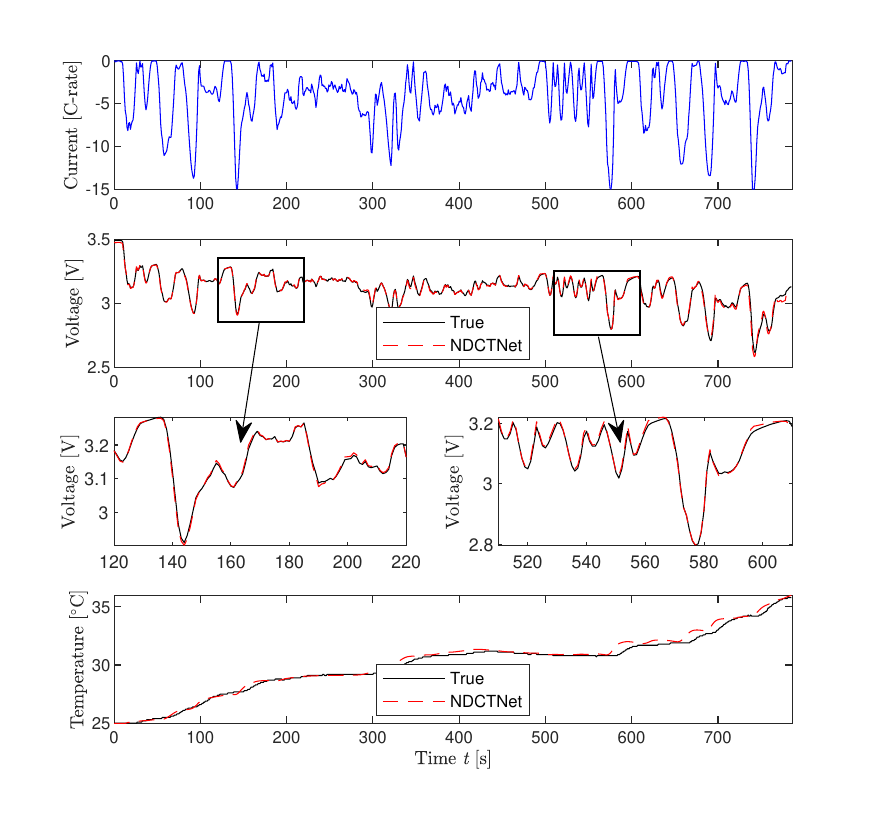}}
    \subfloat[]{
    \centering
    \includegraphics[width = .48\textwidth,trim={1cm .7cm 1cm .9cm},clip]{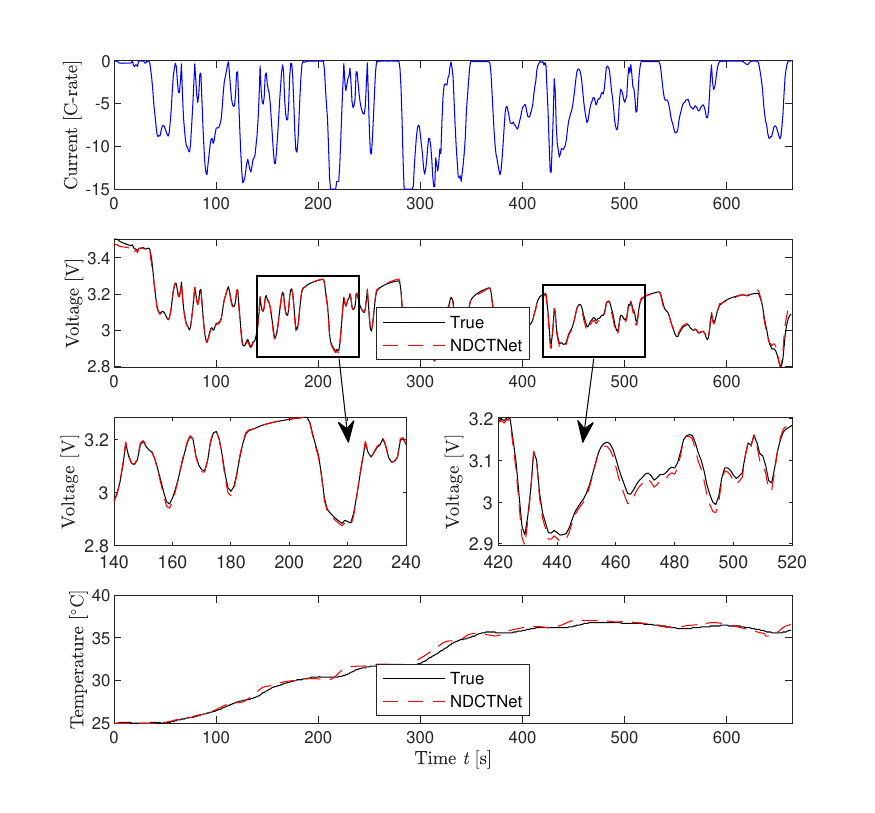}}
    \caption{Validation of the NDCTNet model for the LFP cell based on (a) the modified US06 (0$\sim$15 C) and (b) the modified SC04 (0$\sim$15 C) profiles.}
    \label{Fig: SC04-Model-LFP}
    
\end{figure*}

\begin{figure*}[t!]
\centering
    \subfloat[]{
    \centering
    \includegraphics[width = .48\textwidth,trim={.3cm .6cm 1.32cm .7cm},clip]{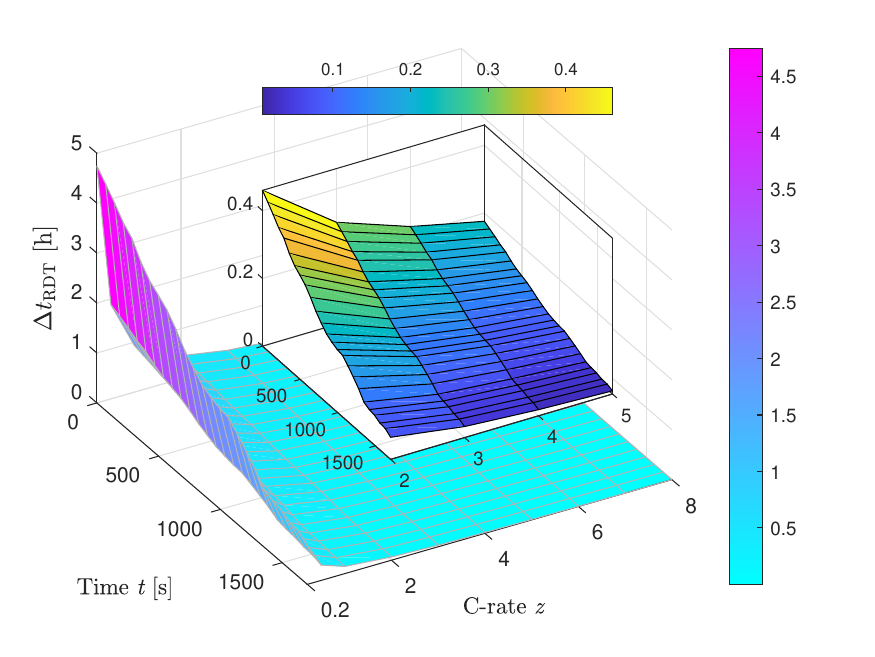}
    \label{Fig: US06-RDTT-NCA-R}}
    \quad
    \subfloat[]{
    \centering
    \includegraphics[width = .48\textwidth,trim={.3cm .6cm 1.32cm .7cm},clip]{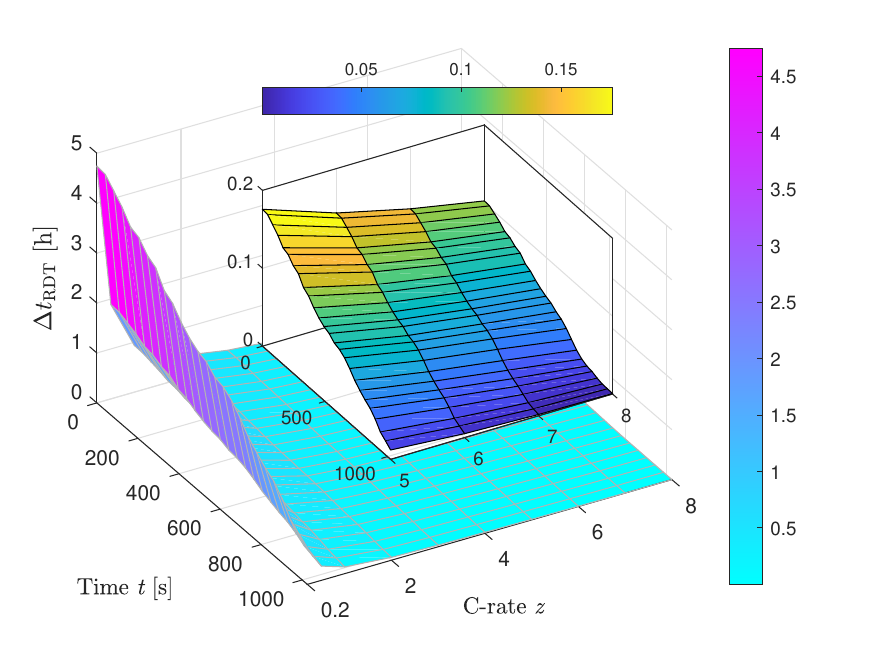}
    \label{Fig: SC04-RDTT-NCA-R}}
    
    \caption{RDT prediction results for the NCA cell under (a) the  modified US06 (0$\sim$8 C) and (b) the  modified SC04 (0$\sim$8 C) testing profiles at $T_{\mathrm{amb}}=25$$^{\circ}\mathrm{C}$.  The inset plots highlight the RDT at high C-rate loads.}
    \label{Fig: US06-RDT-NCA}
\end{figure*}

\begin{figure*}[t!]
\centering
    \subfloat[]{
    \centering
    \includegraphics[width = .48\textwidth,trim={.3cm .6cm 1.32cm .7cm},clip]{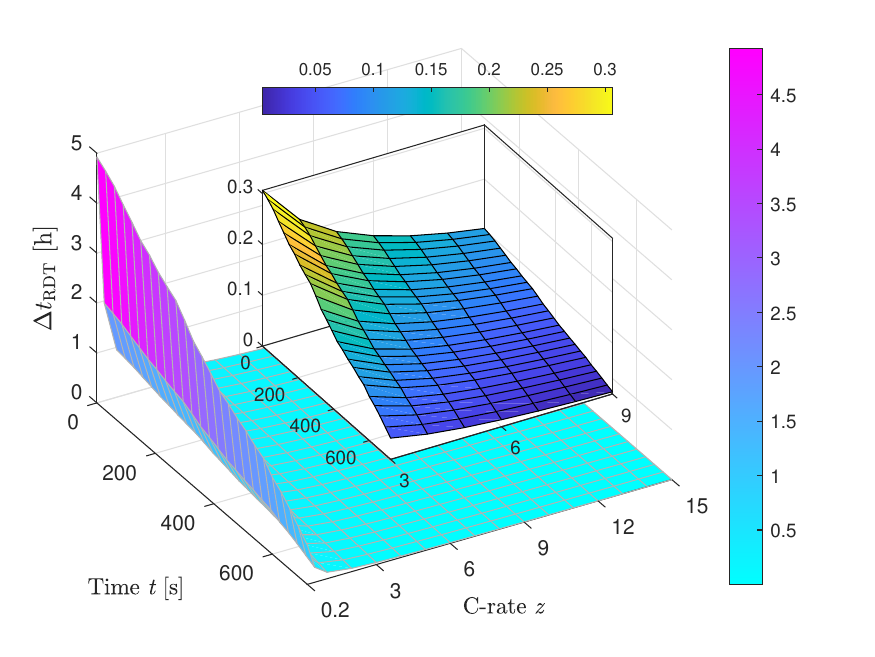}
    \label{Fig: US06-RDTT-LFP}}
    \quad
    \subfloat[]{
    \centering
    \includegraphics[width = .48\textwidth,trim={.3cm .6cm 1.32cm .7cm},clip]{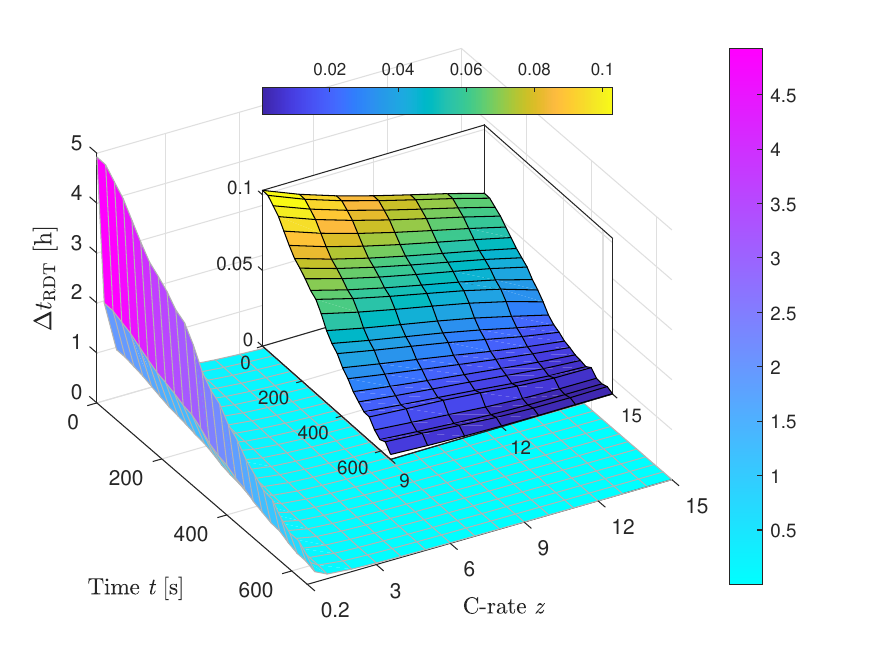}
    \label{Fig: SC04-RDTT-LFP}}
    
    \caption{RDT prediction results for the LFP cell under (a) the  modified US06 (0$\sim$15 C) and (b) the  modified SC04 (0$\sim$15 C) testing profiles at $T_{\mathrm{amb}}=25$$^{\circ}\mathrm{C}$.  The inset plots highlight the RDT at high C-rate loads.}
    \label{Fig: SC04-RDT-LFP}
\end{figure*}

We used the following procedure to construct and validate the NDCTNet model for each of the two cells. 

As the first step, we identified the NDC model and the lumped thermal model for each cell based on the methods in~\cite{Tian:TCST:2021,Lin:JPS:2014}. Then, we moved on to train the FNNs of each NDCTNet model.

\begin{itemize}

\item The NDC model was identified using the parameter identification 1.0 approach in~\cite{Tian:TCST:2021}. The lumped thermal model had been identified for the Samsung NCA cell in~\cite{Biju:AE:2023} and for the A123 LFP cell in~\cite{Lin:JPS:2014}. We made some fine tuning of the parameters to make them fit better with our data.

\item We used the same FNN architecture for $h_V$ and $h_T$ for both cells.  The FNN has two hidden layers with 48 neurons in each hidden layer.  The softplus function is used as the activation function of the hidden layers, and the output layer employs the linear activation function.  The FNNs are configured and trained using Keras, a Python-based deep learning library. The input data normalization was applied to the FNNs.

\item For the NCA cell, we collected the training datasets via constant-current discharging at 0.5/1/3/5/8 C and variable-current discharging based on the modified HWFET/UDDS/WLTC/LA92 profiles (scaled to 0$\sim$8 C)~\cite{EPA}.  The datasets include 22,770   datapoints in total.

\item For the LFP cell, the training datasets were generated by applying constant-current discharging at 0.5/1/2/4/6/8/10/12/15 C and variable-current discharging based on the  modified HWFET/UDDS/WLTC/LA92 profiles (scaled to 0$\sim$15 C).  The datasets contain 23,341 datapoints in total.

\item We use the root-mean-square error (RMSE) as a metric to evaluate the NDCTNet model's prediction performance:
\begin{align*}
\mathrm{RMSE} = \sqrt{\frac{1}{N }\sum_{i=1}^{N } \left(X_{\mathrm{true},i} -X_{\mathrm{NDCTNet},i}  \right)^2},
\end{align*}
where $X_{\mathrm{true},i}$ is the true voltage/temperature at the $i$-th datapoint,  $X_{\mathrm{NDCTNet},i}$ is the predicted voltage/temperature, and $N$ is the total number of datapoints. 

\end{itemize}
Further, we collected extra datasets based on the  modified US06/SC04 profiles, scaled to 0$\sim$8 C for the NCA cell and 0$\sim$15 C for the LFP cell, to test the NDCTNet model after training.  Figures~\ref{Fig: US06-Model-NCA}-\ref{Fig: SC04-Model-LFP} present the validation results, which show excellent fitting performance. Table~\ref{Table: Model Validation} offers a quantitative evaluation  based on the RMSE, indicating that the NDCTNet model exhibits high accuracy over broad current ranges for each cell.

\subsection{Validation of the RDE Prediction Approach}

\begin{figure*}[t!]
\centering
    \subfloat[]{
    \centering
    \includegraphics[width = .48\textwidth,trim={.3cm .6cm 1.5cm .7cm},clip]{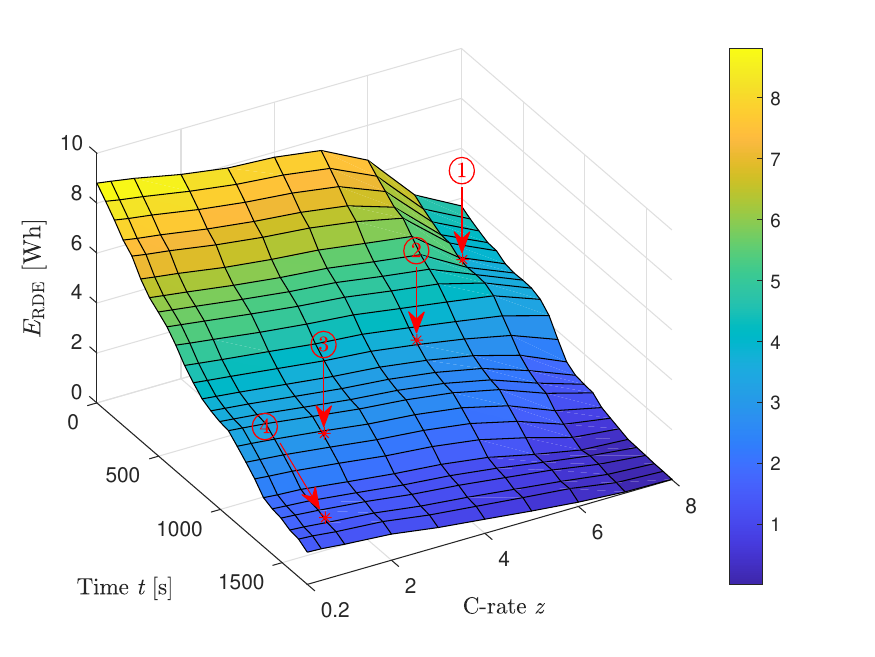}
    \label{Fig: US06-RDET-NCA-R}}
    \quad
    \subfloat[]{
    \centering
    \includegraphics[width = .48\textwidth,trim={.3cm .6cm 1.5cm .7cm},clip]{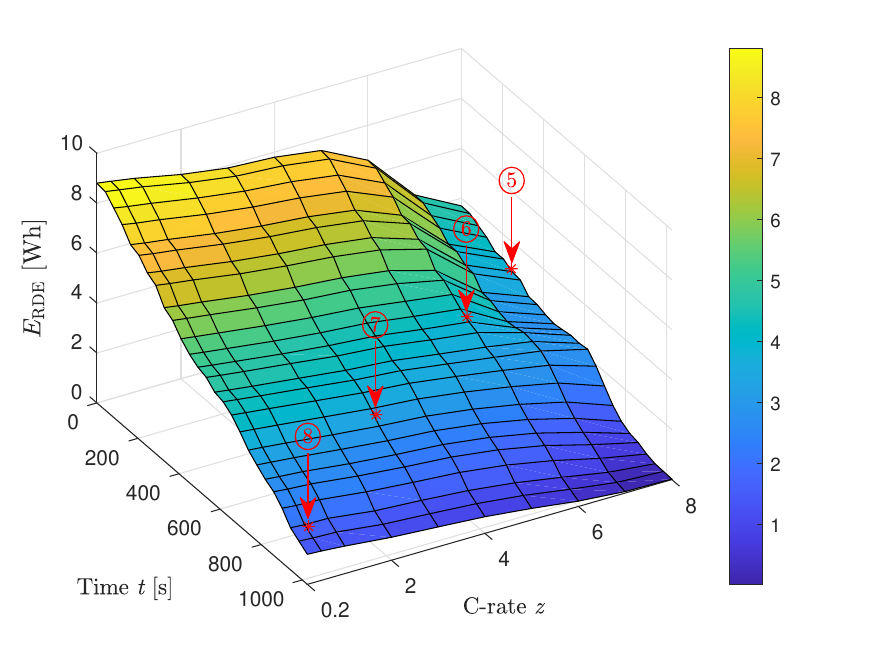}
    \label{Fig: SC04-RDET-NCA-R}}
    
    \caption{RDE prediction results for the NCA cell under (a) the modified US06 (0$\sim$8 C) and (b) the modified SC04 (0$\sim$8 C) testing profiles at $T_{\mathrm{amb}}=25$$^{\circ}\mathrm{C}$. ``$\color{red}*$" denotes the experimental validation datapoint at $\textcircled{\scalebox{0.65}1}\ t= 389\ \mathrm{s}, z=7$, $\textcircled{\scalebox{0.65}2}\ t= 775\ \mathrm{s}, z=5$, $\textcircled{\scalebox{0.65}3}\ t= 1161\ \mathrm{s}, z=2$, $\textcircled{\scalebox{0.65}4}\ t= 1547\ \mathrm{s}, z=1$, $\textcircled{\scalebox{0.65}5}\ t= 243\ \mathrm{s}, z=8$, $\textcircled{\scalebox{0.65}6}\ t= 483\ \mathrm{s}, z=6$, $\textcircled{\scalebox{0.65}7}\ t= 723\ \mathrm{s}, z=3$, $\textcircled{\scalebox{0.65}8}\ t= 963\ \mathrm{s}, z=0.5$.}
    \label{Fig: US06-RDE-NCA}
\end{figure*}

\begin{figure*}[t!]
\centering
    \subfloat[]{
    \centering
    \includegraphics[width = .48\textwidth,trim={.3cm .6cm 1.5cm .7cm},clip]{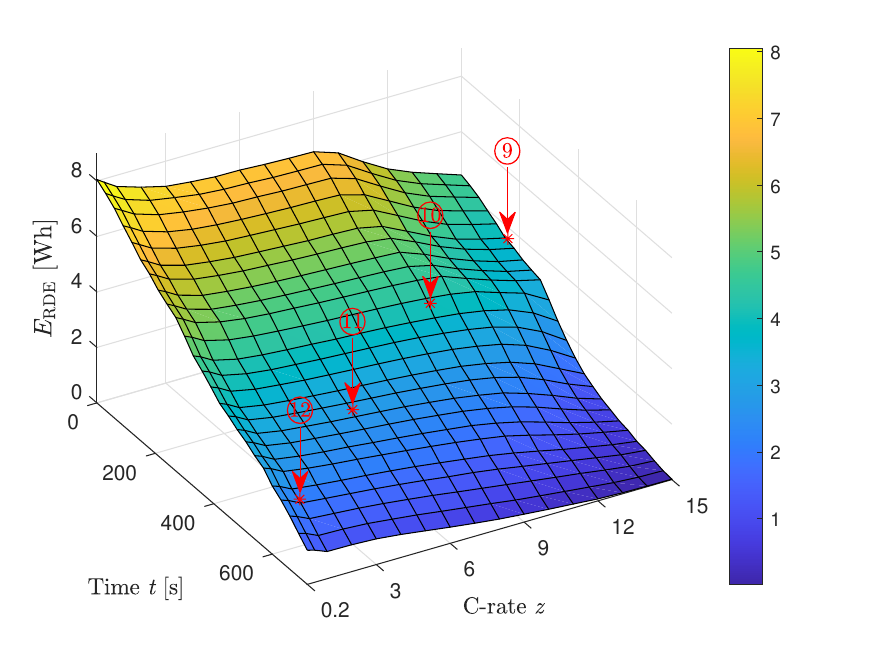}
    \label{Fig: US06-RDET-LFP}}
    \quad
    \subfloat[]{
    \centering
    \includegraphics[width = .48\textwidth,trim={.3cm .6cm 1.5cm .7cm},clip]{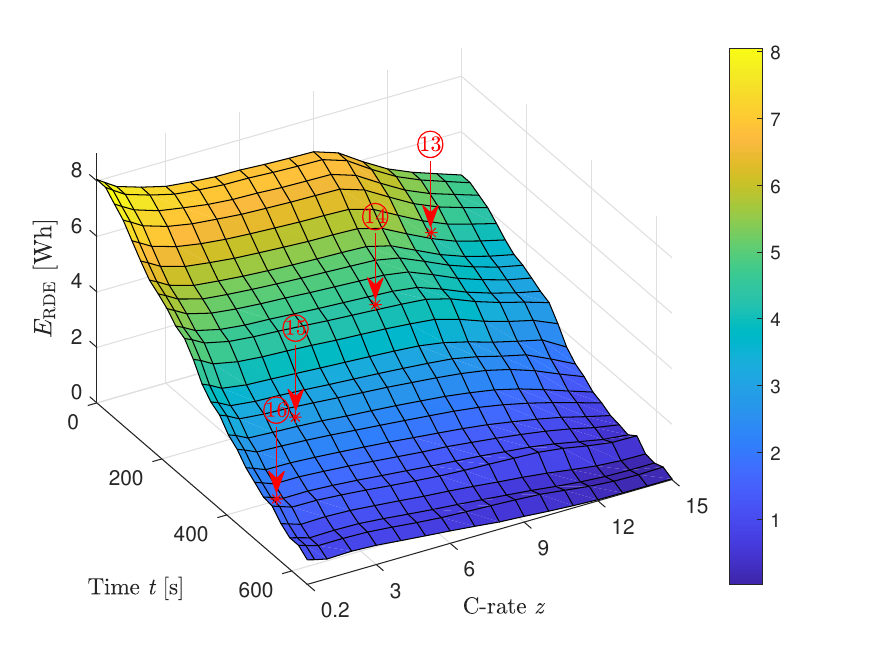}
    \label{Fig: SC04-RDET-LFP}}
    
    \caption{RDE prediction results for the LFP cell under (a) the modified US06 (0$\sim$15 C) and (b) the modified SC04 (0$\sim$15 C) testing profiles at $T_{\mathrm{amb}}=25$$^{\circ}\mathrm{C}$. ``$\color{red}*$" denotes the experimental validation datapoint at $\textcircled{\scalebox{0.65}9}\ t= 159\ \mathrm{s}, z=15$, $\textcircled{\scalebox{0.65}{10}}\ t= 315\ \mathrm{s}, z=10$, $\textcircled{\scalebox{0.65}{11}}\ t= 471\ \mathrm{s}, z=5$, $\textcircled{\scalebox{0.65}{12}}\ t= 627\ \mathrm{s}, z=1$, $\textcircled{\scalebox{0.65}{13}}\ t= 135\ \mathrm{s}, z=12$, $\textcircled{\scalebox{0.65}{14}}\ t= 267\ \mathrm{s}, z=8$, $\textcircled{\scalebox{0.65}{15}}\ t= 399\ \mathrm{s}, z=3$, $\textcircled{\scalebox{0.65}{16}}\ t= 531\ \mathrm{s}, z=0.5$.}
    \label{Fig: SC04-RDE-LFP}
\end{figure*}

\begin{table*}[t!]\centering

 \resizebox{2\columnwidth}{!}{
\begin{tabular}{l l l l l l l l l l l l l}

\toprule
&\makecell[l]{ LiB \\ type} & \makecell[l]{ Current \\ profile} & \makecell[l]{ Time \\ instant \\ $t$ [s]} &\makecell[l]{ C-rate \\ $z$} & \makecell[l]{ True \\ $\Delta t_{\mathrm{RDT}}$ \\{[s]}} & \makecell[l]{ Predicted \\ $\Delta t_{\mathrm{RDT}}$ [s] \\ (proposed \\ method)} & \makecell[l]{Relative \\ $\Delta t_{\mathrm{RDT}}$ \\ error \\ (proposed \\ method)} & \makecell[l]{True \\ $E_{\mathrm{RDE}}$ \\ {[Wh]} } &\makecell[l]{Predicted \\ $E_{\mathrm{RDE}}^{\mathrm{Trad}}$ [Wh] \\ (traditional \\ method)} & \makecell[l] {Relative \\ $E_{\mathrm{RDE}}^{\mathrm{Trad}}$ \\ error \\ (traditional \\ method)} & \makecell[l]{ Predicted \\ $E_{\mathrm{RDE}}$ [Wh] \\ (proposed \\ method)} & \makecell[l] { Relative \\ $E_{\mathrm{RDE}}$ \\ error \\ (proposed \\ method)} \\
\midrule
\makecell[l]{{\color{red} \textcircled{\scalebox{0.65}1}}} & \multirow{8}{*}{NCA}
 & \multirow{4}{*}{\makecell[l]{ Modified\\ US06\\(0$\sim$8 C)}} & 389 & 7 & 221 & 231 & 4.52 \% & 3.64 & 6.27 & 72.25 \% & 3.56 & 2.20 \% \\
\textcircled{\scalebox{0.65}2}& & & 775 & 5 & 279  & 279 & 0 \% & 3.14 & 4.14 & 31.85 \% & 3.11 & 0.96 \% \\
\textcircled{\scalebox{0.65}3}& & & 1161  & 2 & 569 & 574 & 0.88 \% & 2.66 & 3.01 & 13.16 \% & 2.62 & 1.50 \% \\
\textcircled{\scalebox{0.65}4}& & & 1547  & 1 & 601  & 639  & 6.32 \% & 1.47 & 1.58 & 7.48 \% & 1.43 & 2.72 \% \\
\cmidrule(r){1-1}
 \cmidrule(l){3-13}
\makecell[l]{{\color{red} \textcircled{\scalebox{0.65}5}}} & & \multirow{4}{*}{\makecell[l]{ Modified\\ SC04\\(0$\sim$8 C)}} & 243  & 8 & 148  & 154  & 4.05 \% & 2.81  & 6.98  & 148.40 \% & 2.73  &  2.85 \%\\
\textcircled{\scalebox{0.65}6}& & & 483  & 6 & 267  & 276  & 3.37 \% & 3.69 & 4.97 & 34.69 \% & 3.58 & 2.98 \% \\
\textcircled{\scalebox{0.65}7}& & & 723  & 3 & 435  & 445  & 2.30 \% & 3.09 & 3.63 & 17.48 \% & 2.99 & 3.24 \% \\
\textcircled{\scalebox{0.65}8}& & & 963  & 0.5 & 1343  & 1396  & 3.95 \% & 1.63 & 1.66 & 1.84 \% & 1.58 & 3.07 \% \\
\midrule
\makecell[l]{{\color{red} \textcircled{\scalebox{0.65}9}}} & \multirow{12}{*}{LFP}
& \multirow{4}{*}{\makecell[l]{ Modified\\ US06\\(0$\sim$15 C)}} & 159 & 15 & 126  & 120  & 4.76 \% & 3.50  & 6.42  & 83.43 \% & 3.59 & 2.57 \% \\
\textcircled{\scalebox{0.65}{10}} & & & 315  & 10 & 192  & 195  & 1.56 \% & 3.87  & 4.99  & 28.94 \% & 3.84 & 0.78 \% \\
\textcircled{\scalebox{0.65}{11}} & & & 471  & 5 & 269  & 263 & 2.23 \% & 2.72 & 3.47 & 27.57 \% & 2.79 & 2.57 \% \\
\textcircled{\scalebox{0.65}{12}} & & & 627  & 1 & 932  & 894  & 4.08 \% & 1.92 & 2.22 & 15.63 \% & 1.96 & 2.08 \% \\
 \cmidrule(r){1-1}
 \cmidrule(l){3-13}
\makecell[l]{{\color{red} \textcircled{\scalebox{0.65}{13}}}} & & \multirow{4}{*}{\makecell[l]{ Modified\\ SC04\\(0$\sim$15 C)}} & 135  & 12 & 188  & 181  & 3.72 \% & 4.33  & 6.50  & 50.12 \% & 4.39 & 1.39 \% \\
\textcircled{\scalebox{0.65}{14}}& & & 267  & 8 & 253 & 253  & 0 \% & 4.11 & 5.04 & 22.63 \% & 4.15 & 0.97 \% \\
\textcircled{\scalebox{0.65}{15}}& & & 399  & 3 & 426  & 412  & 3.29 \% & 2.63 & 3.02 & 14.83 \% & 2.70 & 2.66 \% \\
\textcircled{\scalebox{0.65}{16}}& & & 531  & 0.5 & 1601  & 1548  & 3.31 \% & 1.68 & 1.77 & 5.36 \% & 1.72 & 2.38 \% \\
 \cmidrule(r){1-1}
 \cmidrule(l){3-13}
\makecell[l]{{\color{red} \textcircled{\scalebox{0.65}{17}}}} & & \multirow{4}{*}{\makecell[l]{Notional \\ eVTOL}} & 50  & 15 & 148  & 141  & 4.73 \% & 4.15  & 7.47  & 80.00 \% & 4.25  & 2.41 \% \\
\textcircled{\scalebox{0.65}{18}}& & & 300  & 10 & 253 &  258 & 1.98 \% & 5.03 & 6.41 & 27.44 \% & 5.17  & 2.78 \% \\
\textcircled{\scalebox{0.65}{19}}& & & 700  & 5 & 398  & 392  & 1.51 \% & 4.01 & 5.05 & 25.94 \% & 4.09  & 2.00 \% \\
\textcircled{\scalebox{0.65}{20}}& & & 900  & 1 & 1745  &  1749 & 0.23 \% & 3.79  & 4.17 & 10.03 \% & 3.81 & 0.53 \% \\
\bottomrule
\end{tabular}
}
\caption{RDE prediction results of the 20 experimental validation datapoints. The preset $T_\mathrm{max}$ limit is triggered at {\color{red} \textcircled{\scalebox{0.65}1} \textcircled{\scalebox{0.65}5} \textcircled{\scalebox{0.65}9} \textcircled{\scalebox{0.65}{13}} \textcircled{\scalebox{0.65}{17}}} .}
\label{Table: RDE prediction}

\end{table*}

Based on the identified NDCTNet model, we continue to validate the proposed RDE prediction approach. First, we simulate the discharging processes using the NDCTNet model and leverage the obtained synthetic data to train the FNNs used in the approach. After the training, we validate the approach by comparing the prediction with the experimental results based on additional discharging experiments.

In the training stage, the setting is as follows.
\begin{itemize}

\item The cut-off limits in the RDE prediction are: $V_{\mathrm{min}}=3\ \mathrm{V}$ and $T_{\mathrm{max}}=50^\circ\mathrm{C}$ for the NCA cell, and $V_{\mathrm{min}}=2.7\ \mathrm{V}$ and $T_{\mathrm{max}}=45^\circ\mathrm{C}$ for the LFP cell. They are chosen based on manufacturer specifications and related literature~\cite{Bandhauer:JES:2011}.

\item All the FNNs used in the RDE prediction for both cells have the same architecture of two hidden layers with each having 48 neurons.  They are also configured and trained using Keras as in Section~\ref{Sec: Exp - NDCTNet}.

\item We construct synthetic training datasets via simulation based on the NDCTNet models. For the NCA cell, the discharging simulation is based on the modified variable-current HWFET/UDDS/WLTC/LA92 profiles (scaled to 0$\sim$8 C). In each discharging case, we branch out to run episodic simulations at every time instant $t$, which discharge the cell fully to $V_\mathrm{min}$ at different constant C-rates ranging between 0.2$\sim$8 C. With the generated data, we determine $\Delta t_\mathrm{RDT}^{V_\mathrm{min}}$ that corresponds to the cell's state $x(t)$ and C-rate $z$. Meanwhile, we compute $E(z,t,\Delta t)$ for each $\Delta t \in \left[0, \Delta t_\mathrm{RDT}^{V_\mathrm{min}}\right]$. These data are utilized to train the FNNs in the RDE prediction approach, with 160,867 datapoints for the $\mathrm{FNN}_\mathrm{RDT}^{V_\mathrm{min}}$ network and 33,068,654 datapoints for the $\mathrm{FNN}_{E}$ network.  The same is done for the LFP cell, with the current range in the simulation adjusted to be 0.2$\sim$15 C.  The training datasets contain 263,195 datapoints for $\mathrm{FNN}_\mathrm{RDT}^{V_\mathrm{min}}$ and 36,439,162 datapoints for $\mathrm{FNN}_{E}$.

\item We introduce a percentage relative error to quantify the prediction accuracy:
\begin{equation*}
     \mathrm{Relative\ error} =\left|\frac{ Y_{\mathrm{true}} - Y_{\mathrm{pred}} }{ Y_{\mathrm{true}} }\right| \times 100\%,
  \end{equation*}
where $Y$ is $\Delta t_{\mathrm{RDT}}$ or $ E_{\mathrm{RDE}}$.
\end{itemize}

Following the training, we apply the approach to the two cells discharged with the modified US06/SC04 profiles for the purpose of validation. For each time instant $t$ during the discharge, we predict the RDE for different C-rate loads between $z=0.2$$\sim$$8$ C for the NCA cell and $z=0.2$$\sim$$15$ C for the LFP cell. Figures~\ref{Fig: US06-RDT-NCA}-\ref{Fig: SC04-RDT-LFP} show the predicted RDT as part of the RDE prediction, in which the inset plots highlight the RDT at high C-rates. Figures~\ref{Fig: US06-RDE-NCA}-\ref{Fig: SC04-RDE-LFP} illustrate the RDE prediction results. We observe that both the RDT and RDE decrease as C-rate increases for both cells due to the rate-capacity effect. Moreover, the predicted RDE drops sharply at very high C-rates ($z = 6$$\sim$$8$ C for the NCA cell and $z = 10$$\sim$$15$ C for the LFP cell). This is because the preset $T_\mathrm{max}$ limit is triggered, which further limits $E_\mathrm{RDE}$.

We further perform experiments to assess the RDE prediction accuracy for each cell. In each experiment, we discharge the cell using a variable-current profile for some time $t$ and then switch to constant-current discharging at a specified C-rate $z$ so as to determine the actual $E_\mathrm{RDE}$. For each cell, both the modified US06/SC04 profiles are used, and for each profile, four different $(t,z)$ pairs are considered. This thus involves experiments for 16 cases in total, as are shown in Figures~\ref{Fig: US06-RDE-NCA}-\ref{Fig: SC04-RDE-LFP}. Table~\ref{Table: RDE prediction} presents a quantitative comparison for all these cases. It is seen that the relative error in the RDE prediction comes less than 3\% in general. Besides, when $T_\mathrm{max}$ is triggered, $E_\mathrm{RDE}$ becomes less and more difficult to be predicted. Table~\ref{Table: RDE prediction} shows the cases when this happens, in all of which accurate prediction is still observed.

Table~\ref{Table: Freq comp} compares the computational time of the proposed ML approach with the model forward simulation. Each prediction runs the two methods at all $z = 0.2/0.5/1/2/3/4/5/6/7/8$ C for the NCA cell and all $z=0.2/0.5/1/2/3/4/5/6/7/8/9/10/11/12/13/14/15$ C for the LFP cell to obtain the RDE corresponding to each specified C-rate. The model forward simulation calculates the RDE by running the NDCTNet model until the cut-off limit is reached. Our ML approach needs only 0.3$\sim$0.4 s for prediction at each time. By contrast, the model forward simulation needs an average time of about 25$\sim$50 s, while demanding a much larger actual time at the beginning of a discharging process.

We further compare our results with a traditional RDE defined as in~\cite{Quade:B&S:2023}:
\begin{align*}
    E_{\mathrm{RDE}}^{\mathrm{Trad}} = Q_{a}\int_{\mathrm{SoC}=0}^{\mathrm{SoC}(t)}  U_{\mathrm{OCV}}(\mathrm{SoC}) d\mathrm{SoC},
\end{align*}
where $Q_{a}$ and $U_{\mathrm{OCV}}$ are the cell's charge capacity in Ah and OCV, respectively. As shown in Table~\ref{Table: RDE prediction}, the prediction of $E_{\mathrm{RDE}}^{\mathrm{Trad}}$ at low C-rates shows a decent accuracy relative to the truth. However, its accuracy declines greatly as the C-rate increases and becomes almost trivial at very high C-rates. This is because $E_{\mathrm{RDE}}^{\mathrm{Trad}}$, by definition, fails to capture the effects of high C-rates on RDE.



\begin{figure*}[t!]

\centering
    \subfloat[]{
    \centering
    \includegraphics[width = .48\textwidth,trim={.8cm .3cm 1cm .7cm},clip]{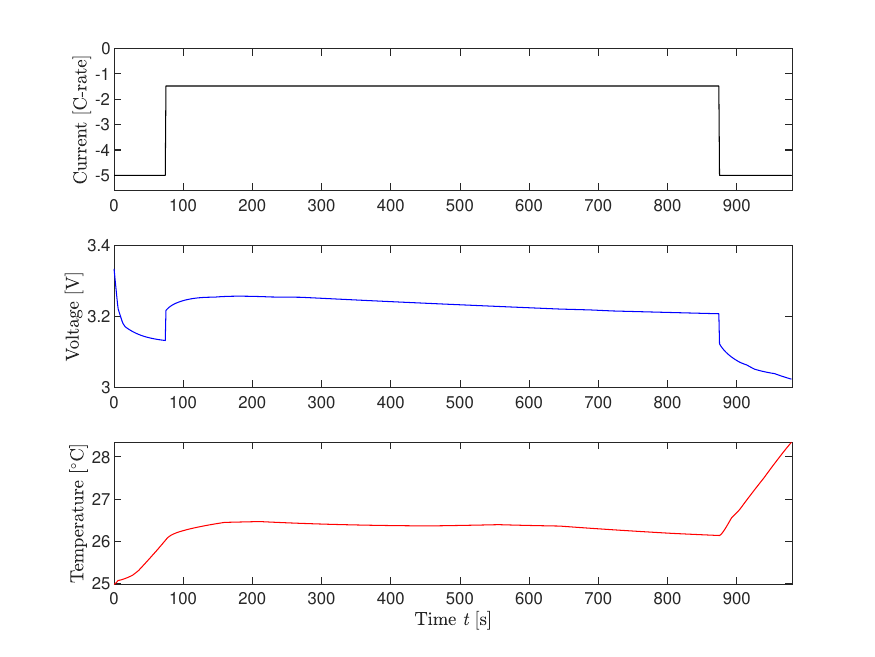}
    \label{Fig: eVTOL-NDCTNet}}
    \quad
    \subfloat[]{
    \centering
    \includegraphics[width = .48\textwidth,trim={.3cm .6cm 1.5cm .7cm},clip]{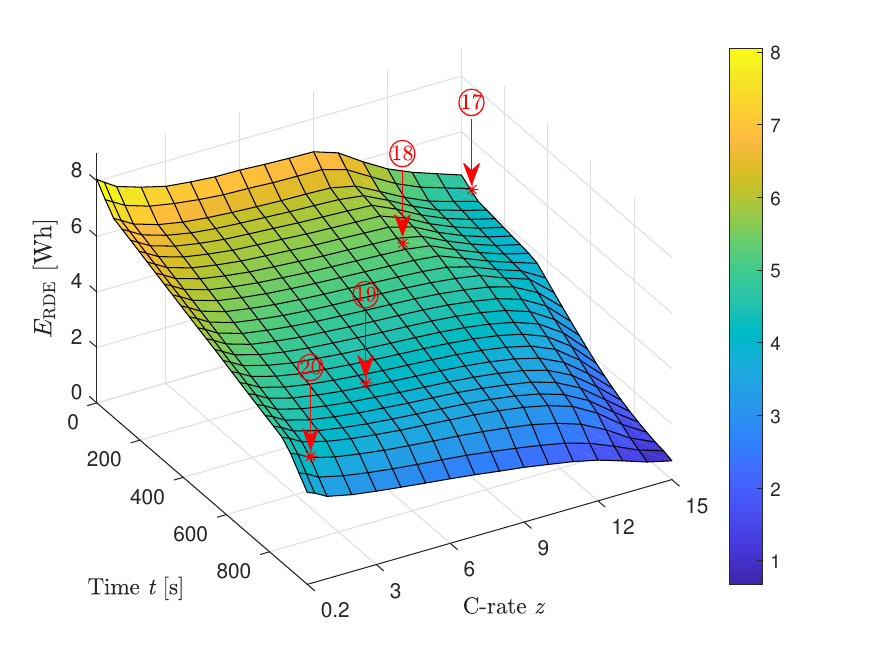}
    \label{Fig: eVTOL-RDET-LFP}}
    
    \caption{Demonstration of the RDE prediction approach under a notional eVTOL profile~\cite{Bills:SD:2023}. (a) Current load, voltage and temperature profiles; (b) RDE prediction results for the LFP cell at $T_{\mathrm{amb}}=25$$^{\circ}\mathrm{C}$. ``$\color{red}*$" denotes the experimental validation datapoint at $\textcircled{\scalebox{0.65}{17}}\ t= 50\ \mathrm{s}, z=15$, $\textcircled{\scalebox{0.65}{18}}\ t= 300\ \mathrm{s}, z=10$, $\textcircled{\scalebox{0.65}{19}}\ t= 700\ \mathrm{s}, z=5$, $\textcircled{\scalebox{0.65}{20}}\ t= 900\ \mathrm{s}, z=1$.}
    \label{Fig: eVTOL-RDE-LFP}
    
\end{figure*}

Moving forward, we present a validation example based on electric vertical take-off and landing (eVTOL) aircraft. The eVTOL technology holds a promise for transforming future air mobility. Differing from electric cars and various other battery-powered applications, it requires high C-rate discharging in take-off and landing~\cite{Bills:SD:2023}. However, as mentioned in Section~\ref{sec: Introduction}, RDE prediction for eVTOL has been studied rarely, despite its importance for safety and performance,  and our proposed approach addresses this gap. In our experimental validation, we simulate a notional flight with discharging rates of 5 C, 1.48 C, and 5 C during the take-off, cruise, and landing phases, respectively, using the aforementioned LFP cell. Figure~\ref{Fig: eVTOL-NDCTNet} illustrates the cell's current, voltage, and temperature profiles. Figure~\ref{Fig: eVTOL-RDET-LFP} presents the RDE prediction for C-rates ranging from $z=0.2$$\sim$$15$ C at various times during the flight. To assess prediction accuracy, we conducted discharging tests at four different time instants with varying C-rates, as shown in Figure~\ref{Fig: eVTOL-RDET-LFP}. Table~\ref{Table: RDE prediction} provides a quantitative comparison, with relative errors of less than 3\% across all cases.


\begin{table}[t!]

\centering
 \resizebox{.95\columnwidth}{!}{
\begin{tabular}{ c c c }
\toprule
\multirow{3}{*}{Method} & \multicolumn{2}{c} {Average RDE computation time} \\ \cmidrule(l){2-3}
 & NCA & LFP \\
& ($z=0.2$$\sim$$8$ C) & ($z=0.2$$\sim$$15$ C) \\
\midrule
Model forward simulation & 25 s & 50 s \\
Proposed ML approach  & 0.3 s & 0.4 s \\
\bottomrule
\end{tabular}}
\caption{Comparison of the computational speed of the RDE prediction methods for the NCA and LFP cell.}
\label{Table: Freq comp}

\end{table}

\section{Discussion}
We provide the following remarks based on the results in Sections~\ref{Sec: RDE definition}-\ref{Sec: Exp validation}.

\begin{itemize}
    \item Predictive accuracy. The proposed ML-based RDE prediction approach demonstrates a high accuracy in experimental validation. This is due to two main reasons. First, the proposed ML architecture, which includes multiple learning modules, is well-founded on technical rationale. Second, we generate large amounts of synthetic but high-fidelity data using the NDCTNet model. This makes it possible to push the training accuracy to the best possible extent for the proposed approach.
    
    \item Computational efficiency. The proposed approach can update the RDE with high computational speed, by leveraging ML to predict the RDE directly using the cell's present state. Compared to it, existing RDE methods have to run a LiB model forward from the present state until the cutoff limit is reached, causing heavy and even prohibitive computation.

    \item Prospective applications. The proposed approach uniquely enables RDE prediction across a wide range of C-rates, making it particularly suitable for high-power LiB applications. One notable example is eVTOL aircraft, which require discharging at up to 5 C during the takeoff and landing phases, and around 1.5 C during the cruise phase~\cite{Bills:SD:2023}. To our knowledge, no existing methods can accurately predict RDE for eVTOL applications. Furthermore, when applied to lithium-ion battery systems operating at low to medium currents, the proposed approach will hold a promise to improve both the accuracy and computational speed of RDE predictions.

    \item Potential extensions. The proposed work offers several avenues for further research. While we have validated our approach on NCA and LFP cells, it would be valuable to apply it to other types of LiBs, such as NMC and LCO. Determining the optimal FNN architectures within our approach remains an intriguing question. We have empirically adjusted the architectures in the study to achieve the best performance, but an automated architecture search could be beneficial. Moreover, our approach using FNNs does not quantify uncertainty, though uncertainty quantification could be important for assessing the confidence levels associated with the RDE prediction. To address this, we could explore the use of probabilistic ML techniques, such as Bayesian neural networks~\cite{Ghahramani:Nature:2015}.

\end{itemize}

\section{Conclusion}
RDE prediction plays an important role in managing the operation of LiBs for high safety and performance. In pursuit of this topic, we consider a new and open question: how to predict the RDE of a LiB cell if the discharging currents span from low to high C-rates? The question has emerged in some applications like electric aircraft but is non-trivial to deal with. This is because high currents will significantly impact the cell's electro-thermal dynamics to make the RDE more difficult to identify. To address the question, we define the RDE as a C-rate-dependent quantity. With the new definition, we develop an ML approach that uses FNNs to grasp the mapping from the cell's present state to the RDE. The approach determines the cell's RDT due to voltage and temperature limits and then uses it along with the cell's state to find out the RDE. To enable the training of the approach, we also develop a hybrid physics-ML model to capture the cell's electro-thermal dynamics. The experimental validation on NCA and LFP cells shows the high prediction accuracy achieved by the proposed approach. Further, the approach is tractable for training and computationally efficient to run. It can find prospective use in different applications ranging from electric cars, heavy-duty vehicles, and aircraft to grid-scale energy storage.

\appendix
\renewcommand{\theequation}{A.\arabic{equation}}
\section*{Appendix}
We derive the explicit formulas for $\phi$ under constant $z$ and $T_\mathrm{amb}$.

Letting $x(t)$ be the initial condition, and considering that $x$ is decomposed into $x_1 = \begin{bmatrix} V_b & V_s & V_1\end{bmatrix}^\top$ and $x_2 = \begin{bmatrix} T_\mathrm{core} & T_\mathrm{surf} \end{bmatrix}^\top$. Based on~(\ref{Eqn: NDC-Dynamic}), the solution of $x_1$ is given by
\begin{align*}
    x_1(t+\Delta t) = e^{A_\mathrm{NDC} \Delta t} x_1(t) + \int_t^{t +\Delta t} e^{A_\mathrm{NDC} (t + \Delta t-\tau)}B_\mathrm{NDC}I(\tau)d\tau,
\end{align*}
which, when $I = zc_o$, becomes
\begin{align}\label{Eqn: V(t)}
    x_1(t+\Delta t) = e^{A_\mathrm{NDC} \Delta t} x_1(t) + \int_t^{t +\Delta t} e^{A_\mathrm{NDC} (t + \Delta t-\tau)}d\tau \cdot B_\mathrm{NDC}zc_o.
\end{align}
It can be easily proven that $A_\mathrm{NDC}$ has three distinct eigenvalues $\lambda_i$ for $i=1,2,3$. By the Cayley-Hamilton theorem, we have
\begin{align} \label{Eqn: MatrixExp}
    e^{A_\mathrm{NDC} \Delta t} = [\Psi_1^{-1}\psi_1(\Delta t)] \oplus A_\mathrm{NDC},
\end{align}
where
\begin{align*}
\Psi_1 =
\begin{bmatrix}
    1 & \lambda_1 & \lambda_1^2 \\
    1 & \lambda_2 & \lambda_2^2 \\
    1 & \lambda_3 & \lambda_3^2 
\end{bmatrix},
\hspace{2mm}
\psi_1(\Delta t) = 
\begin{bmatrix}
    e^{\lambda_1 \Delta t} & e^{\lambda_2 \Delta t} & e^{\lambda_3 \Delta t}
\end{bmatrix}^\top,
\end{align*}
and the operator $\oplus$ denotes 
\begin{align*}
    a \oplus A = \sum_{i=1}^n a_i A^{i-1},
\end{align*}
for $a \in \mathbb{R}^{n\times1}$ and $A \in \mathbb{R}^{n\times n}$. Given (\ref{Eqn: MatrixExp}), the integral in (\ref{Eqn: V(t)}) can be expressed as
\begin{align}\label{Eqn: intMatrixExp}
    \nonumber
    \int_t^{t +\Delta t} e^{A_\mathrm{NDC} (t + \Delta t - \tau)}&d\tau
    \nonumber
    = \int_t^{t +\Delta t} [\Psi_1^{-1}\psi_1(t + \Delta t - \tau)] \oplus A_\mathrm{NDC} d\tau \\
    \nonumber
    &=  [\Psi_1^{-1}\int_t^{t +\Delta t}\psi_1(t + \Delta t - \tau)d\tau] \oplus A_\mathrm{NDC} \\
    &= [\Psi_1^{-1}\int_0^{\Delta t}\psi_1(\zeta)d\zeta] \oplus A_\mathrm{NDC}  \\
    \nonumber
    &=  [\Psi_1^{-1}(\Bar{\psi}_1(\Delta t)-\Bar{\psi}_1(0))] \oplus A_\mathrm{NDC} ,
\end{align}
where $\Bar{\psi}_1(\alpha) = \begin{bmatrix} e^{\lambda_1 \alpha}/\lambda_1 & e^{\lambda_2 \alpha}/\lambda_2 & e^{\lambda_3 \alpha}/\lambda_3 \end{bmatrix}^\top$. Inserting (\ref{Eqn: MatrixExp}) and (\ref{Eqn: intMatrixExp}) into (\ref{Eqn: V(t)}), we have
\begin{align*}
    x_1(t+\Delta t) &=  [\Psi_1^{-1}\psi_1(\Delta t)] \oplus A_\mathrm{NDC} \cdot x_1(t) \\ &\quad +  [\Psi_1^{-1}(\Bar{\psi}_1(\Delta t)-\Bar{\psi}_1(0))] \oplus A_\mathrm{NDC} \cdot B_\mathrm{NDC}zc_o .
\end{align*}
Similarly, we can obtain
\begin{align*}
    x_2(t+\Delta t) &=  [\Psi_2^{-1}\psi_2(\Delta t)] \oplus A_\mathrm{therm} \cdot x_2(t) \\ &\quad +  [\Psi_2^{-1}(\Bar{\psi}_2(\Delta t)-\Bar{\psi}_2(0))] \oplus A_\mathrm{therm} \cdot B_\mathrm{therm} u .
\end{align*}
where $u = \begin{bmatrix} (zc_o)^2 & T_\mathrm{amb}\end{bmatrix}^\top$,
\begin{align*}
\Psi_2 =
\begin{bmatrix}
    1 & \lambda_4 \\
    1 & \lambda_5 \\
\end{bmatrix},
\hspace{2mm}
    \psi_2(\Delta t) = 
\begin{bmatrix}
    e^{\lambda_4 \Delta t} & e^{\lambda_5 \Delta t}
\end{bmatrix}^\top,
\end{align*}
and $\Bar{\psi}_2(\alpha) = \begin{bmatrix} e^{\lambda_4 \alpha}/\lambda_4 & e^{\lambda_5 \alpha}/\lambda_5 \end{bmatrix}^\top$. Here, $\lambda_i$ for $i = 4,5$ are the two distinct eigenvalues of $A_\mathrm{therm}$.









\section*{Acknowledgement}
This work was supported in part by the U.S. National Science Foundation under Award CMMI-1847651 and by the U.S. Department of Energy under Award DE-EE0010404.

\balance
\bibliographystyle{elsarticle-num}
\bibliography{BIB-Energy-Pred}

\end{document}